\documentclass[conference,onecolumn]{IEEEtran}
\usepackage{graphicx}
\usepackage{amssymb,amsmath,latexsym,amsfonts}
\usepackage{epsfig}
\usepackage{comment}
\usepackage{subfigure}
\usepackage{color}

\newtheorem{definition}{Definition}
\newtheorem{theorem}{Theorem}
\newtheorem{lemma}{Lemma}
\def\done{\hspace*{\fill} \rule{1.8mm}{2.5mm}}

\begin{document}
%
\title{Network Non-neutrality Debate: An Economic Analysis}

\author{\IEEEauthorblockN{Eitan Altman, Arnaud Legout, and Yuedong Xu}
\IEEEauthorblockA{INRIA Sophia Antipolis,
2004 Route des Lucioles, Sophia Antipolis, France\\
Email: \{eitan.altman,arnaud.legout\}@inria.fr, yuedong.xu@gmail.com}}
\maketitle

\begin{abstract}
This paper studies economic utilities and quality of service
(QoS) in a two-sided non-neutral market where
Internet service providers (ISPs) charge content providers (CPs)
for the content delivery. We propose new models
that involve a CP, an ISP, end users and advertisers.
The CP may have either a \emph{subscription} revenue model (charging
end users) or an \emph{advertisement} revenue model (charging advertisers).
We formulate the interactions between the ISP and the CP as a noncooperative game
for the former and an optimization problem for the latter.
Our analysis shows that the revenue model of the CP plays
a significant role in a non-neutral Internet. With the \emph{subscription} model,
both the ISP and the CP receive better (or worse) utilities as well as QoS in the presence of the side payment at the same time.
With the \emph{advertisement} model,
the side payment impedes the CP from investing on its contents.

\end{abstract}

\section{Introduction}

Network neutrality, one of the foundations of Internet, is commonly admitted that ISPs
must not discriminate traffic in order to favor
specific content providers \cite{HW}. However, the principle of
network neutrality has been challenged recently. The main reason is
that new broadband applications cause huge amount of traffic without generating direct
revenues for ISPs. Hence, ISPs want to get additional revenues from
CPs that are not directly connected to them. For
instance, a residential ISP might want to charge Youtube in order to give
a premium quality of service to Youtube traffic. This kind of monetary
flows, which violate the principle of network neutrality, are called
\textit{two-sided payment}. We use the term \textit{side payment}
to name the money charged by ISPs from CPs exclusively.

On the one hand, the opponents of network neutrality
argue that it does not give any incentive for ISPs to
invest in the infrastructure. This incentive issue is even more
severe in two cases: the one of tier-one ISPs that support a high
load, but do not get any revenue from CPs; and the one of 3G wireless
networks that need to invest a huge amount of money to purchase spectrum.
On the other hand, advocates of network neutrality claim
that violating it using side payment will lead to unbalanced revenues
among ISPs and CPs, thus a market instability.

Recent work addressed the problem of network neutrality from
various perspectives
\cite{Walrand09,Eitan1,Eitan2,Economides08,Claudia10,Choi08,ToN1:Ma,ToN2:Ma}.
Among these work, \cite{Walrand09,Eitan1,Eitan2} are the closest to
our work.  Musacchio et al. \cite{Walrand09} compare one-sided and
two-sided pricing of ISPs. However, they only investigate an example
where the joint investments of CPs and ISPs bring revenue from
advertisers to CPs. In \cite{Eitan1}, the authors show how side
payment is harmful for all the parties involved such as ISP and CP.
Altman et al. in \cite{Eitan2} present an interesting bargaining framework
to decide how much the ISP should charge the CP.
However, their models might give a biased conclusion by overlooking
the end users' sensitivity towards the prices of the CP and the ISP.

In this paper, we unravel the conflicts of the side payment in a more general context.
We consider a simplified market composed of one ISP, one CP,
some advertisers, and a large number of end users. The ISP
charges end-users based on their usage and sets their QoS level
according to the price paid. The CP can either
have a \emph{subscription} based or an
\emph{advertisement} based revenue model. For the subscription based
revenue model, the CP gets its revenue from the subscription paid by
end-users. End-users adapt their demand of
content based on the price paid to the ISP and the CP.
For the advertisement based revenue model, the CP gets its revenue
from advertisers. End users adapt the demand according to
the price paid to the ISP and the investment of CP on its contents.

Our work differs from related work \cite{Walrand09,Eitan1,Eitan2} by: i) incorporating the QoS provided
by the ISP, ii) studying different revenue models of the CP, and iii) introducing
the \emph{relative price sensitivity} of end users in the subscription model.
Especially, in the subscription model, the \emph{relative price sensitivity} decides whether the side payment
is beneficial (or harmful) to the ISP and the CP. Our finding contradicts the previous
work (e.g. \cite{Eitan1}) that argues that the side payment is harmful for all parties involved.
In the advertisement model, the ability of CP's investment to attract the traffic of end users plays a key role.
It determines whether the side payment is profitable for the ISP and the CP.
Our main contributions are the following.

\begin{itemize}

\item We present new features in the mathematical modeling that include the QoS, the relative price sensitivity of end-users,
and the CP's revenues.

\item We model the price competition between the ISP and the \emph{subscription} based CP as a noncooperative game
and analyze the properties of the Nash equilibrium. The interaction between the ISP and the
\emph{advertisement} based CP is modeled as an optimization problem. The optimal investment
is shown to be a decreasing function of the side payment (from the CP to the ISP).


\item We utilize bargaining games to analyze how the side payment is determined.

\end{itemize}

The rest of this work is organized as follows. In section \ref{sec:model}, we model the economic behaviors
of ISP, CP, advertisers and end users. Section \ref{sec:subscription} and \ref{sec:advertisement}
study the impact of side payment and how the ISP and the CP reach a consensus of side payment.
Section \ref{sec:evaluation} presents
numerical study to validate our claims. Section \ref{sec:conclusion} concludes this paper.

\section{Basic Model}
\label{sec:model}

In this section, we first introduce the revenue models of the ISP and the CP. Then,
we formulate a game problem and an optimization problem for the selfish ISP
and the CP. Finally, we describe the bargaining games in a two-sided market.

\subsection{Revenue Models}
\label{sec:2.1}

We consider a simplified networking market with four economic entities, namely the advertisers,
the CP, the ISP and end users.
All the end users can access the contents of the CP \emph{only} through the network
infrastructure provided by the ISP.
The ISP collects subscription fees from end users. It sets two market
parameters $(p^s, q)$ where $p^s$ is the non-negative price of per-unit of demand,
and $q$ is the QoS measure (e.g. delay, loss or rejection probability).
End users can decide whether to connect to the ISP or not, or how much traffic
they will request, depending on the bandwidth price and the QoS. The CP usually has two
revenue models, the \emph{user subscription} and the \emph{advertisement} from clicks
of users. These two models, though sometimes coexisting with each other, are studied
separately in this work for clarity.
The CP and the ISP interact with
each other in a way that depends on the CP's revenue models.
In the subscription based model, the CP competes with the ISP by charging users a price $p^c$ per-unit of
content within a finite time. End users respond to $p^s$, $p^c$ and $q$ by setting their demands elastically.
Though $p^c$ has a different unit as $p^s$, it can be mapped from the price
per content into the price per bps (i.e. dividing the price of a content by its size
in a finite time). The price $p^c$ not only can stand for a financial disutility, but also
can represent the combination of this disutility together with a cost per quality. Thus
a higher price may be associated with some better quality
(this quality would stand for parameters different than the
parameter $q$ which we introduce later).
Without loss of generality, $p^s$ and $p^c$ can be positive or 0. For the \emph{advertisement}
based model, instead of charging users directly, the CP attracts users' clicks on online
advertisements. The more traffic demands end users generate, the higher the CP's revenue.

\begin{figure}[htb]
\centering
\includegraphics[width=4.0in]{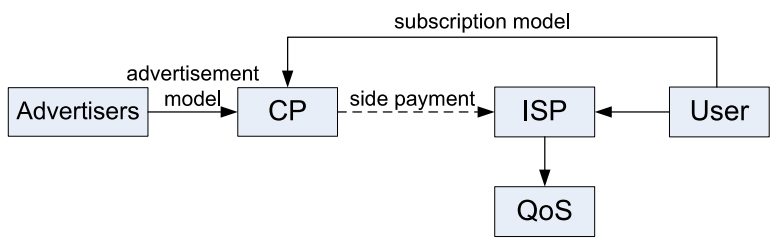}
\caption{Money flow of a non-neutral network.}
\label{fig:moneyflow}
\end{figure}

To better understand network neutrality and non-neutrality, we describe the monetary flows among
different components. The arrows in Figure \ref{fig:moneyflow} represent the recipients of money.
A ``neutral network'' does not allow an ISP to charge a CP for which it is not a direct provider
for sending information to this ISP's users. On the contrary, monetary flow from a CP to an ISP appears when
``network neutrality'' is violated.
The ISP may charge the CP an additional amount of
money that we denote by $f(D) = p^tD$ where $p^t$ is the price of per-unit of demand.
We denote by $\delta\in [0, 1]$ the tax rate of this transferred revenue imposed by the regulator or
the government to the ISP.


We present market demand functions for the \emph{subscription} and the \emph{advertisement} based revenue models.

\subsubsection{Subscription model}
Let us define the average demand of all users by $D$ that has
\begin{equation}
D(p^s,p^c,q) = \max\{0, D_0 - \alpha(p^s + \rho p^c) + \beta q\}, \label{demandfunc}
\end{equation}
\noindent where $D_0$, $\alpha$, $\beta$ and $\rho$ are all positive constants. The parameter $D_0$ reflects the total potential
demand of users. The parameters $\alpha$ and $\beta$ denote the responsiveness of demand to the price and the QoS
level of the ISP. The physical meaning of \eqref{demandfunc} can be interpreted in this way. When the prices of the ISP and the CP
increase (resp. decrease), the demand decreases (resp. increases). If the QoS of the ISP is improved, the demand from users
increases correspondingly. The parameter $\rho$ represents the relative sensitivity of $p^c$ to $p^s$.
We deliberately set different sensitivities of end users to the prices of the ISP and the CP because $p^c$ and $p^s$
refers to different type of disutilities.
If $\rho = 1$, the prices of the ISP and the CP are regarded as having the same effect on $D$. When $\rho > 1$, users are more
sensitive to the change of $p^c$ than $p^s$.
The positive prices $p^s$ and $p^c$ can not be arbitrarily high. They must guarantee a nonnegative demand $D$.

We denote $\mathcal{S}_{CP}$ to be the
pricing strategy of the CP that has $\mathcal{S_{CP}} = \{p^c: p^c \geq 0\}$.
The utility (or revenue equivalently) of the CP is expressed as
{\small
\begin{equation}
U_{cp} = (p^c - p^t)D(p^s,p^c,q). \label{cputility_model}
\end{equation}
}
\noindent Note that the variable $D(p^s,p^c,q)$ is interchangeable with $D$ all the time.
Next, we present the utility of the ISP with QoS consideration. We assume that the pricing strategy of the ISP is defined
by $\mathcal{S}_{ISP} =\{(p^s,q): p^s\geq 0; 0<q\leq q_{max}\}$.
To sustain a certain QoS level of users, the ISP has to pay the costs for operating the backbone,
the last-mile access, and the upgrade of the network, etc. Let $u(D,q)$ be the amount
of bandwidth consumed by users that depends on the demand $D$ and the QoS level $q$.
We assume that $u(D,q)$ is a positive, convex and
strictly increasing function in the 2-tuple $(D,q)$. This is reasonable because a larger demand or higher QoS usually
requires a larger bandwidth of the ISP.
We now present a natural QoS metric as the expected delay \footnote{The QoS metric can be the functions of packet loss rate or expected delay etc.}.
The expected delay is computed by the Kleinrock function that
corresponds to the delay of M/M/1 queue with FIFO discipline or
M/G/1 queue under processor sharing \cite{ITC03:Rachid}.
Similar to \cite{ITC03:Rachid}, instead of using the actual delay, we consider
the reciprocal of its square root, $q = \frac{1}{\sqrt{Delay}} = \sqrt{u(D,q) - D}$.
Thus, the cost $C(D,q)$ can be expressed as $C(D,q) = p_r u(D,q) = p_rD + p_rq^2$, where
$p^r$ the price of per-unit of bandwidth invested by the ISP.
Therefore, the cost of the ISP is denoted by $C(D,q) = p^r u(D,q)$. The utility of the ISP is defined
as the difference between revenue and cost:
\begin{equation}
U_{isp} = (p^s-p^r) D(p^s,p^c,q) + (1-\delta)p^tD(p^s,p^c,q) - p^rq^2. \label{eq2.1.1.5_isputility}
\end{equation}

\subsubsection{Advertisement model}
\label{subsec:advertisement}
Nowadays, a small proportion of CPs like Rapidshare
and IPTV providers get their income from end users. Most of other CPs
provide contents for free, but collect revenues from advertisers. The demand from users
is transformed into attentions such as clicks or browsing of online advertisements.
To attract more eyeballs, a CP needs to invest money on its contents, incurring a cost $c$. The investment
improves the potential aggregate demand $D_0$ in return. Let $D_0(c)$ be a concave and strictly increasing
function of cost $c$. With abuse of notations we denote the strategy of the CP by
$\mathcal{S_{CP}} = \{c: c>0\}$. Hence, the demand to the CP and the ISP is written as
\begin{equation}
D = D_0(c) - \alpha p^s + \beta q. \label{eq2.1.2.1_demand3}
\end{equation}
The utility of the ISP is the same as that in \eqref{eq2.1.1.5_isputility}. Next, we describe the
economic interaction between advertisers and the CP. There are $M$ advertisers
interested in the CP, each of which has a fixed budget $B$ in a given time interval (e.g., daily,
weekly or monthly). An advertiser also has a valuation $v$ to declare its maximum willingness to pay
for each attention. The valuation $v$ is a random variable in the range $[0, \overline{v}]$.
Suppose that $v$ is characterized by probability density function (PDF) $x(v)$ and cumulative
distribution function (CDF) $X(v)$. We assume that the valuations of all advertisers
are independent and identically distributed (i.i.d). Let $p^a$ be the price of per attention charged by the CP.
We denote by $D^a(p^a)$ the demand of attentions from advertisers to the CP. Therefore, $D^a$ can be expressed
as \cite{netecon10:liu}
\begin{equation}
D^a = MB\cdot\textrm{ Prob}(v\geq p^a)/p^a = MB\cdot (1-X(p^a))/p^a. \label{eq2.1.2.2_demand4}
\end{equation}
When the CP increases $p^a$, the advertisers will reduce their purchase of attentions. It is also easy to see
that the revenue of advertising, $p^a\cdot D^a$, decreases with regard to $p^a$ either.
However, the attentions that the CP can provide is upper bounded by the demand of users through the ISP.
Then, we can rewrite $D^a$ as that in \cite{netecon10:liu} by
\begin{equation}
D^a = \min \{D, \;\; MB\cdot (1-X(p^a))/p^a\}. \label{eq2.1.2.3_demand5}
\end{equation}
Correspondingly, subtracting investment from revenue, we obtain the utility of the CP by
\begin{equation}
U_{cp} = (p^a - p^t)D^a - c. \label{eq2.1.2.4_CPutility1}
\end{equation}

\begin{lemma}
\label{lemma:normal_property1}
The optimal demand $D^a$ is a strictly decreasing function of $p^a$
if the pdf $x(v)$ is nonzero in $(0, \overline{v})$.
\end{lemma}
\noindent \textbf{Proof:} According to \eqref{eq2.1.2.4_CPutility1}, there has $X(p^a+\epsilon) > X(p^a)$
where $\epsilon$ is an arbitrarily small positive constant. This is because $x(v) > 0$ for $v\in\{0, \overline{v}\}$.
Hence, $D^a$ is a strictly decreasing function of $p^a$. \done

\begin{figure}[htb]
\centering
\includegraphics[width=2.2in]{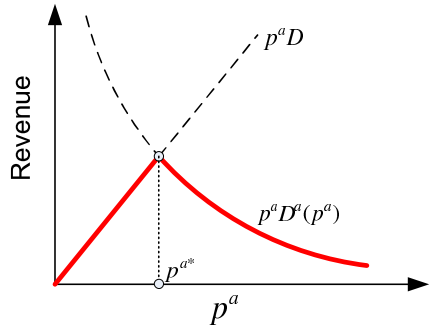}
\caption{Revenue of the CP with regard to $p^a$}
\label{fig:revenue_cp}
\end{figure}

To illustrate how the optimal $p^a$ is found, we follow \cite{netecon10:liu} by drawing figure \ref{fig:revenue_cp}. The X-coordinate
denotes the price $p^a$ and the Y-coordinate denotes the CP's revenue. The curve in red represents the revenue
of the CP coming from advertisers. The optimal price $p^{a*}$ is obtained
at $D = MB\cdot (1-X(p^{a*}))/p^{a*}$. Here, we denote a function $y(\cdot)$ such that $p^{a*} = y(D)$.
According to the demand curve of attentions, $y(\cdot)$ is a decreasing function of $D$.  The utility
of the CP is a function of the demand $D$ and the cost $c$, i.e. $U_{cp} = y(D)\cdot D - p^tD - c$.

\subsection{Problem Formulation}

Two economic entities, the ISP and the CP, want to maximize their utilities.
With the subscription model, the strategy profile of the ISP
is to set the 2-tuple $(p^s, q)$ and that of the CP is to set $p^c$.
This is actually a game problem in which the ISP and the CP compete by setting their prices.
The ISP's QoS is tunable. Thus, we call this game ``QoS Agile Price competition''.
With the advertisement model, the strategy of the ISP is still the price paid by end users,
while that of the CP is to determine the investment level $c$. The ISP and the CP maximize
their own utilities selfishly, but do not compete with each other. We name this maximization
as ``Strategic Pricing and Investment''.


\begin{definition} \label{def_game1}
\emph{\emph{\textbf{\emph{QoS Agile Price Competition}}}}
In the subscription model, the CP charges users based on their traffic demands.
If the Nash equilibrium $NE^{(1)} = \{p^{s*},p^{c*},q^{*}\}$ exists, it can be expressed as
\begin{eqnarray}
(\textbf{G1})\;\;
U_{isp}({p^{s*}, p^{c*}, q^{*}}) &{=}& \max_{\{p^s,q\}\in \mathcal{S}_{ISP}}\!\!\! U_{isp}(p^s, p^{c*}, q),\label{eq:game1_1}\\
U_{cp}(p^{s*}, p^{c*}, q^{*}) &{=}& \max_{p^c\in \mathcal{S}_{CP}} \! U_{cp}(p^{s*}, p^c, q^{*}).\label{eq:game1_2}
\end{eqnarray}
\end{definition}

\begin{definition} \label{def_game2}
\emph{\emph{\textbf{\emph{Strategic Pricing and Investment}}}}
In the advertisement model, the ISP sets $(p^s, q)$ and the CP sets $c$ to optimize their individual utilities.
If there exists an equilibrium $\{p^{s*},q^{*}, c^*\}$, it can be solved by
\begin{eqnarray}
(\textbf{G2})\;\;
U_{isp}({p^{s*}, q^{*}, c^{*}}) &{=}& \max_{\{p^s,q\}\in \mathcal{S}_{ISP}}\!\!\! U_{isp}(p^s, q, c^{*}),\label{eq:game2_1}\\
U_{cp}(p^{s*}, q^{*}, c^{*}) &{=}& \max_{c\in \mathcal{S}_{CP}} \! U_{cp}(p^{s*}, q^{*}, c).\label{eq:game2_2}
\end{eqnarray}
\end{definition}

\subsection{Bargaining Game}

The side payment serves as a fixed parameter in the above two problems.
A subsequent and important problem is how large the side payment should
be. When the ISP decides the side payment unilaterally,
it might set a very high $p^t$ in order to obtain the best utility.
However, this leads to a paradox when the ISP sets $p^s$ and $p^t$ at the meantime.
With the subscription based model, if the ISP plays a strategy $(p^t, p^s, q)$ and
the CP plays $p^c$, the noncooperative game leads to a zero demand and hence a zero income.
This can be easily verified
by taking the derivative of $U_{isp}$ over $p^t$ (also see \cite{Eitan1}).
In other words, the ISP cannot set
$p^t$ and $p^s$ simultaneously in the price competition. Similarly, with the
advertisement based model, the ISP meets with the same paradox.
There are two possible ways, the Stackelberg game and the
bargaining game, to address this problem. Their basic principle is to let the ISP to choose $p^t$ and $p^s$
asynchronously. In this work, we consider the bargaining game in a market
where the ISP and the CP usually have certain marketing powers.
Our analysis in this work is close to the one presented in \cite{Eitan2}, but comes up with
quite different observations.

We here analyze the bargaining games of the side payments that are played at different
time sequences. The first one, namely pre-bargaining, models the situation that
the bargaining takes place before the problems (\textbf{G1}) or (\textbf{G2}).
The second one,
defined as post-bargaining, models the occurrence of bargaining after
the problems (\textbf{G1}) or (\textbf{G2}).
Let $\gamma \in[0,1]$ be the bargaining power of the ISP over the CP.
They negotiate the transfer price $p^t$ determined by
$p^{t*} = \arg\max_{p^t}\;(U_{isp})^{\gamma}(U_{cp})^{1-\gamma}$.
Since the utilities can only be positive, the optimal $p^t$ maximizes a
virtual utility $U$
\begin{eqnarray}
p^t = \arg\max_{p^t}U = \arg\max_{p^t}\;(1-\gamma) \log U_{cp} + \gamma\log U_{isp} \label{eq:sec2.4.1}.
\end{eqnarray}
\noindent We use \eqref{eq:sec2.4.1} to find $p^t$ as a function of the strategies of the ISP and the CP.

\section{Price Competition of the Subscription Model}
\label{sec:subscription}

In this section, we first investigate how the \emph{relative price sensitivity} influences
the price competition between the ISP and the CP. We then study the choice of
side payment under the framework of bargaining games.


\subsection{Properties of Price Competition}
\label{subsec:subscription}

This subsection investigates the impact of side payment on Nash Equilibrium of the noncooperative game \textbf{G1}.
Before eliciting the main result, we show some basic properties of the subscription based revenue model.
\begin{lemma}
\label{lemma:cputility1}
The utility of the CP, $U_{cp}(p^s,p^c,q)$, in \eqref{cputility_model} is a finite, strictly concave function with regard to (w.r.t.) $p^c$.
\end{lemma}

\noindent \textbf{Proof:} Taking the derivative of $U_{cp}$ over $p^c$, we obtain $\frac{d^2U_{cp}}{d{p^c}^2} = -2\alpha\rho <0$.
The demand $D$ is less than $D_0+\beta q_{max}$ so that $p^s$, $p^c$ are finite. Hence, $U_{cp}(p^s,p^c,q)$ is finite
and strictly concave w.r.t. $p^c$. \done

\noindent Similarly, we draw the following conclusion.
\begin{lemma}
\label{lemma:isputility1}
The utility of the ISP, $U_{isp}(p^s,p^c,q)$, in \eqref{eq2.1.1.5_isputility} is a finite, strictly concave function w.r.t. the 2-tuple $(p^s,q)$
if the market parameters satisfy $4\alpha p^r > \beta^2$.
\end{lemma}
\noindent \textbf{Proof:} We compute the Hessian matrix of $U_{isp}$ by
\begin{eqnarray}
J = \left[ \begin{array}{cc}
\frac{d^2U_{isp}}{d(p^s)^2} & \frac{d^2U_{isp}}{dp^s dq} \\
\frac{d^2U_{isp}}{dqdp^s} & \frac{d^2U_{isp}}{dq^2}
\end{array} \right] = \left[ \begin{array}{cc}
-2\alpha & \beta \\
\beta & -2p^r
\end{array} \right].
\end{eqnarray}
When $J$ is negative definite, there must have $J_{1,1} <0$ and $|J_{2,2}| > 0$, that is, $4\alpha p^r > \beta^2$.
If the Hessian matrix if negative definite, $U_{isp}(p^s,p^c,q)$ is strictly concave with regard to the 2-tuple $(p^s,q)$.\done

\noindent For the \emph{QoS Agile Price Competition}, we summarize our main results as below.
\begin{lemma}
\label{theorem:ne1}
When the ISP and the CP set their strategies selfishly,
\begin{itemize}
\item the Nash equilibrium $(p^{s*}, p^{c*}, q^{*})$ is unique;
\item the QoS level $q^*$ at the NE is influenced by the side payment in the ways:
\begin{itemize}
\item improved QoS with $\rho+\delta < 1$;
\item degraded QoS with $\rho+\delta > 1$;
\item unaffected QoS with $\rho+\delta = 1$
\end{itemize}
if $(p^{s*}, p^{c*}, q^{*})$ satisfy $p^{s*} > 0, p^{c*} > 0$ and $0<q^*< q_{max}$.
\end{itemize}
\end{lemma}
\noindent \textbf{Proof:}
Because $U_{cp}$ and $U_{isp}$ are concave functions, the best responses of the ISP and the CP
can be found either inside the feasible region or at the boundary.
In the beginning, we assume that the NE is not at the boundary.
Then, the derivatives $\frac{dU_{isp}}{dp^s}$,
$\frac{dU_{isp}}{dq}$ and $\frac{dU_{cp}}{dp^c}$ are zero at the NE $(p^{s*}, p^{c*}, q^*)$,
\begin{eqnarray}
\frac{dU_{isp}}{dp^s} &=& D - \alpha(p^s+(1-\delta)p^t-p^r) = 0, \label{eq:sec3.2.2}\\
\frac{dU_{isp}}{dq} &=& \beta (p^s+(1-\delta)p^t-p^r) - 2p^rq = 0, \label{eq:sec3.2.3}\\
\frac{dU_{cp}}{dp^c} &=& D - \alpha\rho (p^c - p^t) = 0, \label{eq:sec3.2.4}
\end{eqnarray}
The best response functions of the ISP and the CP are written as
\begin{eqnarray}
p^s &=& \max\{0, \frac{2p^r(D_0-\alpha\rho p^c-\alpha (1-\delta)p^t+\alpha p^r)+ \beta^2((1-\delta)p^t-p^r)}{4\alpha p^r-\beta^2}\}, \label{eq:sec3.2.4_response1}\\
p^c &=& \max\{0, \frac{2p^r(D_0-\alpha p^s+\alpha\rho p^t)+\beta^2(p^s + (1-\delta)p^t-p^r)}{4\alpha \rho p^r}\} \label{eq:sec3.2.4_response2}
\end{eqnarray}
The above equations yield 
\begin{eqnarray}
q^* &=& \frac{\beta(D_0 - \alpha p^r +  \alpha p^t(1-\rho-\delta))}{6\alpha p^r - \beta^2}, \label{eq:sec3.2.5}\\
p^{c*} &=& \frac{2p^r(D_0 - \alpha p^r +  \alpha p^t(1-\rho-\delta))}{\rho(6\alpha p^r - \beta^2)} + p^t, \label{eq:sec3.2.6}\\
p^{s*} &=& \frac{2p^r(D_0 - \alpha p^r +  \alpha p^t(1-\rho-\delta))}{6\alpha p^r - \beta^2} + p^r - (1-\delta)p^t, \label{eq:sec3.2.7} \\
D^{*} &=& \frac{2p^r\alpha(D_0 - \alpha p^r +  \alpha p^t(1-\rho-\delta))}{6\alpha p^r - \beta^2}, \label{eq:sec3.2.8}
\end{eqnarray}
From \eqref{eq:sec3.2.5}, we can see that the QoS is better (resp. worse) if $\rho+\delta<1$ (resp. $\rho+\delta>1$).
The QoS is unchanged with side payment when $\rho$ equals to 1.

Next, we consider the case that the NE is at the boundary. The NE contains three variables so that there are
many possibilities of hitting the boundary. We will not enumerate all of them because the methods of analysis are almost the same.
Here, we study two examples when either $q^*$ in \eqref{eq:sec3.2.5} or $p^{s*}$ in \eqref{eq:sec3.2.7} are outside of $\mathcal{S}_{isp}$.

Obviously, $q^*$ cannot be less than 0. This is because the demand at the NE is non-zero. If $q^*$ in \eqref{eq:sec3.2.5} is larger than $q_{max}$,
$\frac{dU_{isp}}{dq}$ is positive at $q = q_{max}$ when $\frac{dU_{isp}}{dp^s}$ and $\frac{dU_{cp}}{dp^c}$ are zero (i.e. $p^s$ and $p^c$ are
at the equilibrium). Thus, \textbf{G1} is equivalent to the game with a fixed QoS provision. The NE prices $(p^{s*}, p^{c*})$ are characterized by
\begin{eqnarray}
p^{c*} &=& \frac{D_0+\beta q_{max} -\alpha p^r+\alpha(1-\rho-\delta)p^t}{3\alpha \rho} + p^t, \nonumber\\
p^{s*} &=& \frac{D_0+\beta q_{max} -\alpha p^r+\alpha(1-\rho-\delta)p^t}{3\alpha} + p^r - (1-\delta)p^t, \nonumber
\end{eqnarray}
given $p^{c*}$ and $p^{s*}$ are positive.

If $p^{s*}$ in \eqref{eq:sec3.2.7} is negative,
$\frac{dU_{isp}}{dp^s}$ is negative at $p^s = 0$ when $\frac{dU_{isp}}{dq}$ and $\frac{dU_{cp}}{dp^c}$ are zero
The best strategy of the ISP is to set $p^{s*} = 0$ and $q^* = \frac{\beta((1-\delta)p^t-p^r)}{2p^r}$. The demand function turns into $D = D_0 - \alpha\rho p^c + \beta q^{*}$. Thus, the best strategy of the CP is
\begin{eqnarray}
p^{c*} = \frac{2p^r(D_0+\alpha \rho p^t) + \beta^2((1-\delta)p^t-p^r)}{4\alpha \rho p^r}. \label{eq:sec3.2.8_cpbestresponse_boundary}
\end{eqnarray}
Submitting \eqref{eq:sec3.2.8_cpbestresponse_boundary} to \eqref{eq:sec3.2.4_response1}, we validate that the best response of the ISP is $p^{s*} = 0$.

By enumerating all the boundary conditions, we can always find that the NE is unique. When the NE is not at the boundary, the QoS is better with side payment
from the CP to the ISP if $\rho + \delta < 1$, and vice versa. \done

Lemma \ref{theorem:ne1} means that that the QoS provision of the ISP is influenced by the side payment.
We interpret the results by considering $\rho$ and $\delta$ separately. When users are indifferent to the
price set by the ISP and that by the CP (i.e., $\rho = 1$), a positive tax rate $\delta$ leads to the degradation of $q$ in the presence
of side payment. Next, we let $\delta$ be 0 and investigate the impact of $\rho$. If users are more sensitive to the price
of the ISP (i.e. $\rho<1$), the side payment is an incentive of the ISP to improve its QoS. Otherwise, charging side payment leads
to an even poorer QoS of the ISP. According to \eqref{eq:sec3.2.5}$\sim$\eqref{eq:sec3.2.8},
if users are more sensitive to the CP's price,
a good strategy of the ISP is to share its revenue with the CP so that the latter sets a lower subscription fee.

\subsection{Bargaining of Side Payment}
\label{bargain_subs}
To highlight the bargaining of side payment, we make the following two simplifications: i)
the tax ratio $\delta$ is 0, and ii) $p^t$ can be positive, zero or negative.
%
%
%
We let $\delta=0$ because it turns out to have the similar effect as $\rho$.
The above analysis has shown that a negative $p^t$ might benefit both the ISP and the CP in some
situations. Hence, we relax the feasible region of $p^t$ in the bargaining game.

\noindent \textbf{Pre-bargaining:} In the pre-bargaining, $p^t$ is chosen based on
the NE of the ISP and the CP. The equations \eqref{eq:sec3.2.5}$\sim$
\eqref{eq:sec3.2.7} yield the expression of $U$
\begin{eqnarray}
U = 4\log\big(D_0 - \alpha p^r +  \alpha p^t(1-\rho)\big) + \textrm{ constant }. \noindent
\end{eqnarray}
The utility $U$ is increasing or decreasing in $p^t$ depending on the sign of $(1-\rho)$. If $\rho < 1$,
a positive $p^t$ improves not only the QoS level of the ISP, but also the utilities of the ISP and the CP.
As $p^t$ increases, $p^s$ decreases and $p^c$ increases consequently until $p^s$ hits 0. Hence, in the
pre-bargaining, $p^{s*} = 0$. The prices $p^{t*}$ and $p^{c*}$ are computed by
\begin{eqnarray}
p^{t*} &=& \frac{p^r(4\alpha p^r + 2D_0 - \beta^2)}{4\alpha p^r +2\rho\alpha p^r - \beta^2}. \\
p^{c*} &=& \frac{2p^r(D_0 - \alpha p^r +  \alpha p^{t*}(1-\rho-\delta))}{\rho(6\alpha p^r - \beta^2)} + p^{t*}.
\end{eqnarray}
On the contrary, when $\rho > 1$, a negative $p^t$ benefits both of them. Then, $p^{t*}$ is a negative value such that $p^{c*}$ is 0.
When $\rho=1$, the QoS and the utilities are unaffected by any $p^t$. Also, the selection of $p^t$ is uninfluenced by
the bargaining power $\gamma$.

\noindent \textbf{Post-bargaining:} For the post-bargaining, the ISP and the CP compete for the subscription
of users first, knowing that they will bargain over $p^t$ afterwards \cite{Eitan2}. In brief,
we find $p^t$ as a function of $p^s$, $p^c$ and $q$ first. Then, the ISP and the CP compete with each other
by setting the prices. To solve the maximization in \eqref{eq:sec2.4.1}, we let $\frac{dU}{dp^t}$ be 0 and
obtain
\begin{eqnarray}
p^t = \gamma p^c - (1-\gamma)(p^s-p^r) + (1-\gamma)p^rq^2/D. \label{eq:sec4.2.10}
\end{eqnarray}
Submitting \eqref{eq:sec4.2.10} to $U_{isp}$, we rewrite the ISP's utility by
\begin{eqnarray}
U_{isp} = \gamma\big((p^s+p^c-p^r)(D_0-\alpha(p^s+\rho p^c)+\beta q) - p^rq^2   \big). \label{eq:sec4.2.11}
\end{eqnarray}
The utility of the CP is proportional to that of the ISP, i.e. $\frac{U_{isp}}{\gamma} = \frac{U_{cp}}{1-\gamma}$.
After knowing $p^t$, we compute the derivatives $\frac{dU_{isp}}{dp^s}$, $\frac{dU_{isp}}{dq}$ and $\frac{dU_{cp}}{dp^c}$
by
\begin{eqnarray}
\frac{dU_{isp}}{dp^s} &=& \gamma(D - \alpha(p^s+p^c-p^r)), \label{eq:sec4.2.12}\\
\frac{dU_{isp}}{dq} &=& \gamma(\beta(p^s+p^c-p^r) - 2p^rq), \label{eq:sec4.2.13}\\
\frac{dU_{cp}}{dp^c} &=& (1-\gamma)(D - \alpha\rho(p^s+p^c-p^r)). \label{eq:sec4.2.14}
\end{eqnarray}
The best responses of $U_{isp}$ and $U_{cp}$ will not happen
at the same time unless $\rho=1$ or $p^s+p^c-p^r = 0$. The condition $p^s+p^c-p^r = 0$ does
not hold because it leads to a zero demand $D$ and zero utilities.
When $\rho$ is not 1, only one of \eqref{eq:sec4.2.12}
and \eqref{eq:sec4.2.14} is 0. 
Here, we consider the case $\rho>1$. The utility $U_{cp}$ reaches its maximum upon $D = \alpha\rho(p^s+p^c-p^r)$,
while $U_{isp}$ is still strictly increasing w.r.t. $p^s$. Thus, the ISP
increases $p^s$ until the demand $D$ is 0, which contradicts the condition of a nonzero $D$.
If $D = \alpha(p^s+p^c-p^r)$, $\frac{dU_{cp}}{dp^c}$ is negative and $\frac{dU_{isp}}{dp^s}$ is 0.
Then, the CP decreases $p^c$ until 0 and the ISP sets $p^s$ to achieve its best utility accordingly.
By letting \eqref{eq:sec4.2.14} be 0, we can find $(p^{s},q)$ at the Nash equilibrium
\begin{eqnarray}
q^{*} = \frac{\beta(D_0-\alpha p^r)}{4\alpha  p^r - \beta^2} \;\;\;\textrm{     and     }\;\;\;
p^{s*} = \frac{2p^r(D_0-\alpha p^r)}{4\alpha  p^r - \beta^2} + p^r. \nonumber
\end{eqnarray}
The price of side payment, $p^{t*}$, is thus computed by
\begin{eqnarray}
p^t = -(1-\gamma)\frac{D_0-\alpha p^r}{2\alpha}. \label{eq:sec4.2.17}
\end{eqnarray}
When $\rho =1$, $p^{s*}$ and $p^{c*}$ can be arbitrary values in their feasible region that
satisfy $p^{s*} + p^{c*} = p^r + \frac{2p^r(D_0-\alpha p^r)}{4p^r\alpha-\beta^2}$. Similar result has been shown in \cite{Eitan2}.
The analysis of $\rho < 1$ is omitted here since it can be conducted in the same way.

\section{Price, QoS and Investment settings of the Advertisement Model}
\label{sec:advertisement}

This subsection analyzes how the side payment influences the optimal
strategies of the ISP and the CP with the advertisement model.
The bargaining games are adopted to determine
the amount of side payment. Compared with subscription based model, the advertisement
based model exhibits quite different behaviors.

\subsection{Properties of Advertisement Mode}

In general, the subscription model is limited to file storage CDNs, newspaper corporations,
or some big content owners such as movie producers. Most of content
providers are not able to provide enough unique contents so that they do not charge users,
but make money from online advertisements. In this subsection, we present the general properties
of the advertisement model and a couple of case studies.

\begin{lemma}
\label{lemma:isp_ad}
For any feasible investment $c$ of the CP, there exists a best strategy of the ISP, $(p^{s},q)$.
When $c$ increases, the price and the QoS (i.e. $p^s$ and $q$) become larger.
\end{lemma}
\noindent \textbf{Proof:} According to lemma \ref{lemma:isputility1},
$U_{isp}$ is a strictly concave function of 2-tuple $(p^s,q)$ in the feasible region.
When $c$ is determined, the best response $(p^s,q)$ is derived accordingly.

The two-tuple $(p^{s*},q^*)$ at the NE is solved by
\begin{eqnarray}
p^{s*} &=& \frac{2p^r(D_0(c) + \alpha (1-\delta)p^t - \alpha p^r)}{4\alpha p^r - \beta^2}, \label{eq3.3.1.ps}\\
q^* &=& \frac{\beta(D_0(c) + \alpha (1-\delta)p^t - \alpha p^r)}{4\alpha p^r - \beta^2}, \label{eq3.3.2.qos}
\end{eqnarray}
if $p^{s*} > 0$. Since $D_0(c)$ is an increasing function, we can easily find that $p^s$ and $q$ increase w.r.t. $c$.\done

In \textbf{G2}, the CP and the ISP do not compete with each other. On one hand, the ISP
sets the two-tuple $(p^s,q)$ with the observation of $c$. On the other hand, the CP adjusts $c$ based on $(p^s,q)$.
The investment of the CP brings more demand of end users, which increases the revenues of not only the ISP, but also the CP.
Hence, different from \textbf{G1}, the problem \textbf{G2} is not a game. Instead of studying the Nash equilibrium,
we look into the best responses of the ISP and the CP in \textbf{G2}.

\begin{theorem}
\label{theorem:ne2}
There exists a unique best response, namely $(p^{s*}, q^*, c^*)$, with the advertisement
model if the revenue of the CP, $D\cdot y(D)$,
is a concave function w.r.t. $D\geq 0$.
\end{theorem}
\noindent \textbf{Proof:}
In the proof, we will show that the best response equations of the ISP and the CP
have only one solution.

First, we assume that $(p^{s*}, q^*, c^*)$ is not obtained at
the boundary (i.e. $p^{s*}>0$, $c^*>0$ and $0<q^*< q_{max}$). To get the best response functions
of the ISP, we derive $U_{isp}$ over $p^s$ and $q$,
\begin{eqnarray}
\frac{dU_{isp}}{dp^s} &=& D - \alpha(p^s+(1-\delta)p^t-p^r) =0, \label{eq3.3.3.derivative_ps}\\
\frac{dU_{isp}}{dq} &=& \beta(p^s+(1-\delta)p^t-p^r) - 2p^rq =0, \label{eq3.3.3.derivative_q}
\end{eqnarray}
Submitting \eqref{eq3.3.3.derivative_q} and \eqref{eq2.1.2.1_demand3} to \eqref{eq3.3.3.derivative_ps},
we obtain
\begin{eqnarray}
\frac{dU_{isp}}{dp^s} = \frac{(4\alpha p^r-\beta^2)D -2p^r\alpha(D_0(c) + \alpha (1-\delta)p^t - \alpha p^r)}{2\alpha p^r - \beta^2}. \label{eq3.3.3.derivative_ps2}
\end{eqnarray}
The best response functions of the ISP give rise to the demand function
\begin{eqnarray}
D = \frac{2p^r\alpha(D_0(c) + \alpha (1-\delta)p^t - \alpha p^r)}{4\alpha p^r - \beta^2}. \label{eq3.3.3.demand}
\end{eqnarray}
The derivative of $U_{cp}$ over $c$ is expressed as
\begin{eqnarray}
\frac{dU_{cp}}{dc} = (Dy'(D) + y(D) - p^t)\frac{dD_0(c)}{dc} - 1. \label{eq3.3.4.derivativeofCP}
\end{eqnarray}
Letting $\frac{dU_{cp}}{dc} = 0$, we obtain the following equation
\begin{eqnarray}
\frac{dD_0(c)}{dc} = \frac{1}{Dy'(D) + y(D) - p^t}. \label{eq3.3.4.bestrespCP}
\end{eqnarray}
Given that $D_0(c)$ is a strictly concave function of $c$, $\frac{dD_0(c)}{dc}$ is decreasing
in terms of $c$. The revenue of the CP, $y(D)D$, is increasing and concave so that the expression $\frac{1}{Dy'(D) + y(D) - p^t}$
is a non-decreasing function of $D$. In the demand function \eqref{eq3.3.4.bestrespCP}, the left side is a decreasing function
of $c$, while the right side is non-decreasing w.r.t. $D$. This indicates that $D$
is a non-increasing of $c$. However, the demand of the ISP in \eqref{eq3.3.3.demand} is an increasing function
of $c$. Thus, the intersection of \eqref{eq3.3.3.demand} and \eqref{eq3.3.4.bestrespCP} is the unique fixed point.

If \eqref{eq3.3.3.demand} and \eqref{eq3.3.4.bestrespCP} do not have an intersection in the range $[0,\infty)$,
the best response $(p^{s*}, q^*, c^*)$ might be at the boundary. We consider the cases when each variable
in the set $(p^{s*}, q^*, c^*)$ is negative according to the fixed point equation composed of
\eqref{eq3.3.3.demand} and \eqref{eq3.3.4.bestrespCP}. If $q^*<0$, $p^{s}>0$ and $c^*>0$, the demand at
the equilibrium $D^*$ is negative, which is not true. Hence, we only need to consider two cases
of the fixed point equation from \eqref{eq3.3.3.demand} and \eqref{eq3.3.4.bestrespCP}.
i) $\{p^{s*} < 0, c^* > 0\}$ and ii) $\{p^{s*} > 0, c^* < 0\}$.
If the intersection of \eqref{eq3.3.3.demand} and \eqref{eq3.3.4.bestrespCP} has $p^{s*}<0$ and $c^*>0$,
the demand of the end users is $D = D_0(c) + \beta q = D_0(c) + \frac{\beta^2((1-\delta)p^t-p^r)}{2p^r}$
and $p^s$ is chosen to be 0. Submitting the above expression of $D$ to \eqref{eq3.3.4.bestrespCP},
we obtain a new fixed point equation w.r.t. $c$
\begin{eqnarray}
\frac{dD_0(c)}{dc} = \frac{1}{D_0(c) + \frac{\beta^2((1-\delta)p^t-p^r)}{2p^r}y'(D_0(c) + \frac{\beta^2((1-\delta)p^t-p^r)}{2p^r}) + y(D_0(c) + \frac{\beta^2((1-\delta)p^t-p^r)}{2p^r}) - p^t}. \label{eq3.3.5.bestrespCP}
\end{eqnarray}
Because the left side is a decreasing function of $c$ and the right side is an increasing function of $c$,
the solution in the range $[0, \infty)$ is unique if it exists.
For the case that the intersection of \eqref{eq3.3.3.demand} and \eqref{eq3.3.4.bestrespCP} has $c^*<0$ and $p^{s*}>0$,
the CP sets $c$ to be 0. Then, the aggregate total demand is actually $D_0(0)$.
The equations \eqref{eq3.3.3.demand} and \eqref{eq3.3.3.derivative_ps} yield
\begin{eqnarray}
p^{s*} = \frac{2p^r(D_0(0) + \alpha (1-\delta)p^t - \alpha p^r)}{4\alpha p^r - \beta^2} + p^r - (1-\delta)p^t > 0 \label{eq3.3.4.bestDemandwith0c}
\end{eqnarray}
Note in the case ii), if $c^{*}$ derived from the fixed point equation \eqref{eq3.3.5.bestrespCP} is less than 0,
the CP will set the $c^*$ to be 0. Similarly, in case iii), if $p^{s*}$ in \eqref{eq3.3.4.bestDemandwith0c}
is less than 0, the ISP's best strategy is to let $p^{s*} = 0$.

This completes the proof.
\done

\begin{lemma}
\label{lemma:sidepayment_ad}
The side payment from the CP to the ISP leads to a decreased investment on the contents
when the best strategy $(p^{s*}, q^*, c^*)$ has $p^{s*}>0, q^*>0$ and $c^*>0$.
\end{lemma}
\noindent \textbf{Proof:} Given that $(p^{s*}, q^*, c^*)$ has $p^{s*}>0, q^*>0$ and $c^*>0$,
the best strategy is not at the boundary.
Let $p^t_x$ and $p^t_y$ be two prices of side payment that have $p^t_x < p^t_y$. We will prove
$c_x^* > c_y^*$ by contradiction. Assume that $c_x^* \leq c_y^*$ when $p^t_x < p^t_y$. From \eqref{eq3.3.3.demand}
we obtain $D_x^* < D_y^*$. In the ride side of \eqref{eq3.3.4.bestrespCP}, $Dy'(D) + D$ is decreasing with regard to $D$
so that there has $D_x^*y'(D_x^*) + D_x^* - p^t_x > D_y^*y'(D_y^*) + D_y^* - p^t_y$. Hence, we obtain
$\frac{dD_0(c)}{dc}\mid_{c=c_x^*} < \frac{dD_0(c)}{dc}\mid_{c=c_y^*}$. Given $\frac{dD_0(c)}{dc}$ is a decreasing
function of $c$, there has $c_x^* > c_y^*$, which contradicts to the assumption $c_x^* < c_y^*$.
Therefore, the optimal investment of CP is an decreasing function of the price of side payment. \done



\subsection{Case Study}
In this subsection, we aim to find the best strategies of the ISP and the CP
when the valuation of advertisers follows a uniform distribution or a normal distribution.

Recall that the potential aggregate demand of users,
$D_0(c)$, is strictly increasing and concave w.r.t $c$. When the CP
invests money on contents, $D_0$ becomes larger,
while its growth rate shrinks. Here, we assume a log function of $D_0(c)$,
\begin{eqnarray}
D_0(c) = D_0^0 + K\log(1+c), \label{eq:sec3.4.1}
\end{eqnarray}
where the constant $K$ denotes the ability that the CP's investment brings the demand.
The nonnegative constant $D_0^0$ denotes the potential aggregate demand of end users
when $c$ is zero (the CP only provides free or basic contents).
The utility of the ISP remains unchanged.

\noindent\textbf{Uniform Distribution:} Suppose $v$ follows a uniform distribution
in the range $[0, \overline{v}]$. Then, the CDF $X(p^a)$ is expressed as $\frac{p^a}{\overline{v}}$.
The optimal price $p^a$ is obtained when $D = \frac{MB}{p^a}\cdot (1-\frac{p^a}{\overline{v}})$
in the range $[0, \overline{v}]$ (see subsection \ref{subsec:advertisement}). Alternatively, there has
\begin{eqnarray}
p^a = \frac{MB\overline{v}}{MB+D\overline{v}}. \label{eq:sec3.4.3}
\end{eqnarray}
\noindent The above expressions yield the utility of the CP by
\begin{eqnarray}
U_{cp} = \frac{MB\overline{v}D}{MB+D\overline{v}} - c - p^tD . \label{eq:sec3.4.4}
\end{eqnarray}
\noindent Deriving $U_{cp}$ over $c$, we obtain
\begin{eqnarray}
\frac{dU_{cp}}{dc} = (\frac{(MB)^2\overline{v}}{(MB+D\overline{v})^2} - p^t)\cdot \frac{K}{1+c} -1 . \label{eq:sec3.4.5}
\end{eqnarray}
We let \eqref{eq:sec3.4.5} be 0 and get
\begin{eqnarray}
c = K(\frac{(MB)^2\bar{v}}{(MB + D\bar{v})^2} - p^t) - 1.\label{eq:sec3.4.6}
\end{eqnarray}
The rule of the ISP to decide $(p^s, q)$ is the same as that in the subscription model,
except that the aggregate demand is not a constant, but a function of $c$,
\begin{eqnarray}
c = \exp(\frac{\frac{D}{2p^r\alpha}(4p^r\alpha-\beta^2)-D_0^0+\alpha p^r -(1-\delta)p^t\alpha}{K}) - 1.\label{eq:sec3.4.7}
\end{eqnarray}
Note that \eqref{eq:sec3.4.6} is strictly decreasing and \eqref{eq:sec3.4.7} is strictly increasing. They
constitute a fixed-point equation for the 2-tuple $(D^*,c^*)$.
When $D$ approaches infinity, \eqref{eq:sec3.4.6}
is negative while \eqref{eq:sec3.4.7} is positive. When $D$ is zero, if \eqref{eq:sec3.4.6} is larger than \eqref{eq:sec3.4.7},
there exists a unique fixed-point solution. In this fixed point, the ISP and the CP cannot benefit from changing their
strategy unilaterally. We can solve $c^*$ and $D^*$ numerically using a binary search.
If \eqref{eq:sec3.4.6} is smaller than \eqref{eq:sec3.4.7} when $D$ is 0, the best strategy of the CP
is exactly $D=0$. The physical interpretation is that the increased revenue from advertisers cannot
compensate the investment on the contents. Once $D^{*}$ and $c^*$ are derived, we can solve
$p^{s*}$ and $q^*$ subsequently.
In this fixed-point equation, $p^t$ greatly influences the optimal investment $c^*$. When $p^t$ grows, \eqref{eq:sec3.4.6}
and \eqref{eq:sec3.4.7} decrease at the mean time. The crossing point of two curves,
\eqref{eq:sec3.4.6} and \eqref{eq:sec3.4.7}, may shift toward the direction of smaller $c$.
Intuitively, the contents of the CP become less when the ISP charges a positive $p^t$.

Next, we analyze the best strategies of the ISP and the CP when they are at the boundary.
Note that we compute $p^{s*}$ and $c^*$ numerically through \eqref{eq:sec3.4.6} and \eqref{eq:sec3.4.7}.
They might be negative, thus violating their feasible ranges. If the computed $p^{s*}<0$,
the derivative $\frac{dU_{isp}}{dp^s}$ cannot be 0 with $p^{s}=0$.
When the ISP set its price $p^{s*}$ to be 0, the demand function is rewritten as
\begin{eqnarray}
D = D_0^0 + K\log(1+c) + \beta q = D_0^0 + K\log(1+c) + \frac{\beta((1-\delta)p^t-p^r)}{2p^r}. \label{eq:sec3.4.8_demandfunc1}
\end{eqnarray}
Submitting \eqref{eq:sec3.4.8_demandfunc1} to \eqref{eq:sec3.4.6}, we obtain a new fixed point equation in term of $c$.
The best investment of the CP
is obtained by solving this fixed point equation. Similarly, we consider the case that
$c^*$ computed from \eqref{eq:sec3.4.6} and \eqref{eq:sec3.4.7} is less than 0. Due to the constraint $c\geq 0$, the best strategy of the CP
is to set $c^*=0$. When $c^*$ is 0, the demand to the ISP is expressed as
\begin{eqnarray}
D = D_0^0 -\alpha p^s + \beta q. \label{eq:sec3.4.8_demandfunc2}
\end{eqnarray}
By letting $\frac{dU_{isp}}{dp^s}$ and $\frac{dU_{isp}}{dq}$ be 0, the best strategy of the ISP is
represented as the following
\begin{eqnarray}
p^{s*} &=& \frac{2p^r(D_0+(1-\delta)\alpha p^t -\alpha p^r)}{4p^r\alpha - \beta^2} - (1-\delta)p^t + p^r,\label{eq:sec3.4.9_zeroinvest1}\\
q^{*} &=& \frac{\beta(D_0+(1-\delta)\alpha p^t -\alpha p^r)}{4p^r\alpha - \beta^2}. \label{eq:sec3.4.9_zeroinvest2}
\end{eqnarray}

%
%

\noindent\textbf{Normal Distribution:}
Suppose that $v$ obeys a normal distribution with the mean $\mu > 0$ and the deviation $\sigma^2$.
The valuation $v$ ranges from $-\infty$ to $\infty$ (i.e. $\underline{v} = -\infty$ and $\overline{v}= \infty$)
\footnote{Truncated normal distribution is usually adopted in econometrics. In order to keep mathematical simplicity,
we consider a normal distribution.}. The probability distribution function $x(v)$ is expressed as
\begin{eqnarray}
x(v) = \frac{1}{\sqrt{2\pi\sigma^2}}e^{-\frac{(v-\mu)^2}{2\sigma^2}} \label{eq:sec3.4.20_cdf_adv}
\end{eqnarray}
\noindent The CDF of $v$ is computed by $X(v) =  \int_{-\infty}^v x(t) dt$. Since the optimal price $p^{a*}$ (or $y(D)$ alternatively) is obtained
at $D = MB\cdot (1-X(p^{a*}))/p^{a*}$, we obtain $p^{a*}$ as an implicit function of $D$, (denoted by $p^a$)
\begin{eqnarray}
D = \frac{MB}{\sqrt{2\pi\sigma^2}p^{a*}} \int_{p^{a*}}^{\infty} e^{-\frac{(t-\mu)^2}{2\sigma^2}}dt.\label{eq:sec3.4.21_demand_adv}
\end{eqnarray}
\noindent For simplicity, we rewrite $U_{cp}$ as the following
\begin{eqnarray}
U_{cp} = \frac{MB}{\sqrt{2\pi\sigma^2}} \int_{p^{a*}}^{\infty} e^{-\frac{(t-\mu)^2}{2\sigma^2}}dt - c - p^tD.\label{eq:sec3.4.22_utility_adv}
\end{eqnarray}
Let $(p^{s*}, q^{*}, c^*)$ be a set of best strategies of the ISP and the CP (computed from their best response functions).
To prove the uniqueness of $(p^{s*}, q^{*}, c^*)$, we show that $y(D)D$ is a concave function w.r.t. $D$.
\begin{lemma}
\label{lemma:normal_property2}
The revenue of the CP, $y(D)D$, is a concave function of $D$ when the valuation $v$ follows a normal distribution with the mean $\mu>0$
and the variance $\sigma^2$.
\end{lemma}
\noindent \textbf{Proof:} Denote a function $F(D)=y(D)D$ to be the revenue of the CP. To prove the concavity of $F(D)$,
we need to show $\frac{d^2F(D)}{dD^2} \leq 0$. We derive $F(D)$ over $D$ and obtain
\begin{eqnarray}
\frac{dF(D)}{dD} &=& \frac{dF(D)}{dp^{a*}}\cdot \frac{dp^{a*}}{dD} = \frac{dF(D)}{dp^{a*}}\cdot (\frac{dD}{dp^{a*}})^{-1} \nonumber\\
&=& \frac{e^{-\frac{(p^{a*}-\mu)^2}{2\sigma^2}}(p^{a*})^2}{e^{-\frac{(p^{a*}-\mu)^2}{2\sigma^2}}p^{a*} +
\int_{p^{a*}}^{\infty}e^{-\frac{(p^{a*}-\mu)^2}{2\sigma^2}}dt}.\nonumber
\end{eqnarray}
We next compute the derivative $\frac{d^2F(D)}{dD^2}$,
\begin{eqnarray}
\frac{d^2F(D)}{dD^2} &=& \frac{d(\frac{dF(D)}{dD})}{dD} = \frac{d(\frac{dF(D)}{dD})}{dp^{a*}}\cdot (\frac{dD}{dp^{a*}})^{-1} \nonumber\\
&=&-\frac{\frac{\sqrt{2\pi \sigma^2}}{MB}(p^{a*})^3e^{-\frac{(p^{a*}-\mu)^2}{2\sigma^2}}}{(e^{-\frac{(p^{a*}-\mu)^2}{2\sigma^2}}p^{a*} +
\int_{p^{a*}}^{\infty}e^{-\frac{(p^{a*}-\mu)^2}{2\sigma^2}}dt)^3}
\times \Big(2p^{a*}e^{-\frac{(p^{a*}-\mu)^2}{2\sigma^2}} + (2 - \frac{p^{a*}(p^{a*}-\mu)}{\sigma^2})\int_{p^{a*}}^{\infty}e^{-\frac{(p^{a*}-\mu)^2}{2\sigma^2}}dt\Big).
\label{cprevenue_2order_derivative}
\end{eqnarray}
Because $p^{a*}$ is an implicit function of $D$ in \eqref{eq:sec3.4.21_demand_adv}, $p^{a*}$ is positive. Otherwise, the demand $D$
from the ISP is negative, which is not feasible.
Let us denote functions $\Omega(p^{a*})$ and $\Psi(p^{a*})$ by
\begin{eqnarray}
\Omega(p^{a*}) = 2p^{a*}e^{-\frac{(p^{a*}-\mu)^2}{2\sigma^2}}
 + (2 - \frac{p^{a*}(p^{a*}-\mu)}{\sigma^2})\int_{p^{a*}}^{\infty}e^{-\frac{(p^{a*}-\mu)^2}{2\sigma^2}}dt \nonumber
\end{eqnarray}
and
\begin{eqnarray}
\Psi(p^{a*}) = e^{-\frac{(p^{a*}-\mu)^2}{2\sigma^2}}p^{a*} + \int_{p^{a*}}^{\infty}e^{-\frac{(p^{a*}-\mu)^2}{2\sigma^2}}dt \nonumber
\end{eqnarray}
respectively. Here, we will show that $\Omega(p^{a*})$ and $\Psi(p^{a*})$ are positive for $p^{a*} > 0$.
The derivative of $\Omega(p^{a*})$ over $p^{a*}$ is expressed as
\begin{eqnarray}
\frac{d\Omega}{dp^{a*}} = -\frac{p^{a*}(p^{a*}-\mu)}{\sigma^2}e^{-\frac{(p^{a*}-\mu)^2}{2\sigma^2}}
- \frac{2p^{a*}-\mu}{\sigma^2}\int_{p^{a*}}^{\infty}e^{-\frac{(p^{a*}-\mu)^2}{2\sigma^2}}dt <0.\nonumber
\end{eqnarray}
for any $p^a \geq \mu$. Hence, $\Omega(p^{a*})$ is strictly decreasing for $p^{a*} > \mu$.
When $p^{a*}$ goes to infinity, $\Omega(p^{a*})$ approaches 0, i.e. $\lim_{p^{a*}\rightarrow\infty}\Omega(p^{a*}) = 0$.
Because $\Omega(p^{a*})$ is a strictly decreasing function, we can infer $\Omega(p^{a*}) > 0$ for any positive $p^{a*} > \mu$.
According to the expression of $\Omega(p^{a*})$, it is easy to observe $\Omega(p^{a*}) > 0$ for any $p^{a*}$ that
has $0 < p^{a*}\leq \mu$. Thus, $\Omega(p^{a*})$ is positive for any positive $p^{a*}$. We derive $\Psi(p^{a*})$
over $p^{a*}$ subsequently,
\begin{eqnarray}
\frac{d\Psi}{dp^{a*}} =  -\frac{p^{a*}(p^{a*}-\mu)}{\sigma^2} \cdot e^{-\frac{(p^{a*}-\mu)^2}{2\sigma^2}}.\nonumber
\end{eqnarray}
\noindent From the above expression, we can see that $\Psi(p^{a*})$ is increasing for $p^{a*} < \mu$ and decreasing for
$p^{a*} > \mu$. Note that there have $\Psi(0) > 0$ and $\lim_{p^{a*}\rightarrow\infty}\Psi(p^{a*}) = 0$. We can see
that $\Psi(p^{a*})$ is positive in the range $[0, \infty)$. Because $\Psi(p^{a*})$ and $\Omega(p^{a*})$
are positive for $p^{a*}>0$, $\frac{d^2F(D)}{dD^2} < 0$ so that $F(D)$ is a concave function of $D, (D>0)$. \done

The concavity of $y(D)D$ implies the uniqueness of the best strategies of the ISP and the CP. We then show how to find
$(p^{s*}, q^*, c^*)$ numerically. The analysis of the normal distribution is carried out based on the assumption
that $(p^{s*}, q^*, c^*)$ is not at the boundary first. Then, $(p^{s*}, q^*, c^*)$ is the solution of equations
$\frac{dU_{isp}}{dp^s} = 0$, $\frac{dU_{isp}}{dq} = 0$ and $\frac{dU_{cp}}{dc} = 0$.
According to \eqref{eq:sec3.4.21_demand_adv}, the inverse function $D^{-1}(p^a)$ does not have a close form expression.
Hence, the derivative $\frac{dU_{cp}}{dc}$ contains the variables $c$ and $p^{a*}$,
\begin{eqnarray}
\frac{U_{cp}}{dc} &=& \frac{U_{cp}}{dp^{a*}}\cdot \frac{dp^{a*}}{dD} \cdot \frac{dD}{dc} = \frac{U_{cp}}{dp^{a*}}\cdot
\big(\frac{dD}{dp^{a*}}\big)^{-1} \cdot \frac{dD}{dc} \nonumber\\
&=& \Big(\frac{e^{-\frac{(p^{a*}-\mu)^2}{2\sigma^2}}(p^{a*})^2}{e^{-\frac{(p^{a*}-\mu)^2}{2\sigma^2}}p^{a*} +
\int_{p^{a*}}^{\infty}e^{-\frac{(p^{a*}-\mu)^2}{2\sigma^2}}dt} - p^t\Big)\cdot \frac{K}{1+c} - 1. \label{eq:sec3.4.23_derive_utility_adv0}
\end{eqnarray}
By letting $\frac{dU_{cp}}{dc}$ be 0, we obtain
\begin{eqnarray}
c = K\Big(\frac{e^{-\frac{(p^{a*}-\mu)^2}{2\sigma^2}}(p^{a*})^2}{e^{-\frac{(p^{a*}-\mu)^2}{2\sigma^2}}p^{a*} +
\int_{p^{a*}}^{\infty}e^{-\frac{(p^{a*}-\mu)^2}{2\sigma^2}}dt} - p^t\Big) - 1.\label{eq:sec3.4.23_derive_utility_adv}
\end{eqnarray}
The best response of the ISP with the normal distribution
is the same as that with the uniform distribution.
We submit the demand function in \eqref{eq:sec3.4.21_demand_adv} to \eqref{eq:sec3.4.7} and obtain
\begin{eqnarray}
D &=& \frac{MB}{\sqrt{2\pi\sigma^2}p^{a*}} \int_{p^{a*}}^{\infty} e^{-\frac{(t-\mu)^2}{2\sigma^2}}dt \nonumber\\
&=& \frac{2p^r\alpha\big(D_0^0 + K\log(1+c) - \alpha p^r + (1-\delta)p^t\alpha\big)}{4p^r\alpha-\beta^2}.\label{eq:sec3.4.24_derive_utility_isp}
\end{eqnarray}
The equations \eqref{eq:sec3.4.23_derive_utility_adv} and \eqref{eq:sec3.4.24_derive_utility_isp} form a fixed point equation
of $p^{a*}$ and $c$.

The proof of Lemma \ref{lemma:normal_property2} shows that $c$ an increasing function of $p^{a*}$
in \eqref{eq:sec3.4.23_derive_utility_adv}.
While in \eqref{eq:sec3.4.24_derive_utility_isp} $c$ is a decreasing function of $p^a$.
Therefore, the solution of this fixed point is unique if it is not at the boundary.

Next, we investigate the cases where the best strategies are at the boundary. With the advertisement model, the boundary cases
include i) $\{p^{s*} = 0, c>0\}$, ii) $\{p^{s*} > 0, c=0\}$, and iii) $\{p^{s*}= 0, c=0\}$.
If the fixed point solution computed from \eqref{eq:sec3.4.23_derive_utility_adv} and \eqref{eq:sec3.4.24_derive_utility_isp}
has $p^{s*} < 0$, then the derivative $\frac{dU_{isp}}{dp^s}$ is not zero at $p^s=0$. The demand function $D$ is then expressed as
\begin{eqnarray}
D =\frac{MB}{\sqrt{2\pi\sigma^2}p^{a*}} \int_{p^{a*}}^{\infty} e^{-\frac{(t-\mu)^2}{2\sigma^2}}dt
= D_0^0 + K\log(1+c) + \frac{\beta((1-\delta)p^t-p^r)}{2p^r}.\label{eq:sec3.4.24_derive_utility_isp2}
\end{eqnarray}
Here, \eqref{eq:sec3.4.23_derive_utility_adv} and \eqref{eq:sec3.4.24_derive_utility_isp2} consist of a new fixed point equation.
The optimal strategies of the ISP and the CP is the solution of this fixed point equation.
If the fixed point solution computed from \eqref{eq:sec3.4.23_derive_utility_adv} and \eqref{eq:sec3.4.24_derive_utility_isp}
has $c^* < 0$, then the derivative $\frac{dU_{cp}}{dc}$ is not zero $c = 0$. The best response of the ISP is obtained at $c^*=0$.
We can use \eqref{eq:sec3.4.9_zeroinvest1} and \eqref{eq:sec3.4.9_zeroinvest2} to compute $(p^{s*}, q^*)$.

\subsection{Bargaining of Side Payment}

In the bargaining game of advertisement model, $p^t$ can also be negative.
We assume that the valuation $v$ follows a uniform distribution, and makes the same
simplifications as those in subsection \ref{bargain_subs}.

\noindent\textbf{Pre-bargaining:}
The pre-bargaining method determines $p^t$ by maximizing the virtual utility $U$,
\begin{eqnarray}
p^t = \arg\max_{p^t}U = \arg\max_{p^t}\;(1-\gamma)\log(U_{cp}(p^{s*},q^*,c^*)) + \gamma\log(U_{isp}(p^{s*},q^*,c^*))\label{eq:sec4.3.1}\nonumber
\end{eqnarray}
\noindent where the superscript $^*$ denotes the value of a variable at the equilibrium.
In the beginning, we assume that $p^{t*}$ does not cause $p^{s*}$, $c^*$ or $q^*$ outside of their feasible ranges.
Submitting \eqref{eq:sec3.4.4} to \eqref{eq:sec4.3.1}, we obtain
\begin{eqnarray}
p^t &=& \arg\max_{p^t}(1-\gamma) \log \big(\frac{MB\bar{v}D^*}{MB + \bar{v}D^*} - c^* - p^tD^*\big) + \gamma\log\big((p^{s*}+p^t-p^r)D^* - p^r{q^*}^2\big)\label{eq:sec4.3.2} \nonumber\\
&=& \arg\max_{p^t}(1-\gamma) \log \big(\frac{MB\bar{v}D^*}{MB + \bar{v}D^*} - c^* - p^tD^*\big) + 2\gamma \log D^* + \textrm{ constant } \nonumber\\
&=& \arg\max_{p^t}(1-\gamma) \log \big(\frac{MB\bar{v}D^*}{MB + \bar{v}D^*} - K(\frac{M^2B^2\bar{v}}{(MB+\bar{v}D^*)^2}-p^t) +1 - p^tD^*\big) \nonumber\\
& & + 2\gamma \log D^* + \textrm{ constant }.
\end{eqnarray}

Recall that in the pre-bargaining of the subscription model, $p^{t*}$ is independent of the bargaining power $\gamma$.
When the ISP charges the CP $p^t$ for per-unit of demand, either $p^{s*}$ or $p^{c*}$ is zero, depending on $\rho$.
Different from that of the subscription model, the optimal side payment $p^{t*}$ relies on the bargaining parameter $\gamma$.

If $p^{t*}$ causes $c^*<0$ or $p^{s*}<0$, we need to replace the $U_{isp}$ and $U_{cp}$ by the corresponding expressions
in which the best strategies of the ISP and the CP are at the boundary (see subsection \ref{subsec:advertisement}). Using the same
method, we can find the optimal side payment.


\noindent\textbf{Post-bargaining:}

In the post-bargaining of advertisement model, the bargaining of $p^t$ happens after the ISP and the CP select their best policies.
They negotiate $p^t$ to maximize the virtual utility $U$,
\begin{eqnarray}
p^t = \arg\max_{p^t}U = \arg\max_{p^t}\;\;\;(1-\gamma) \log \big((y(D)-p^t)D-c\big) + \gamma\log\big((p^s+p^t-p^r)D - p^rq^2\big) \label{eq:sec4.4.1}.
\end{eqnarray}
Deriving $U$ over $p^t$ and letting the derivative be 0, we obtain
\begin{eqnarray}
p^t = \gamma y(D) - \frac{\gamma c}{D} - (1-\gamma)(p^s-p^r) + (1-\gamma)\frac{p^rq^2}{D} \label{eq:sec4.4.2_pt_postbargain}.
\end{eqnarray}
\noindent Submitting \eqref{eq:sec4.4.2_pt_postbargain} to $U_{isp}$, we have
\begin{eqnarray}
U_{isp} = \gamma \big((y(D)+p^s-p^r)D - c - p^rq^2\big) \label{eq:sec4.4.3}
\end{eqnarray}
and $U_{cp} = \frac{1-\gamma}{\gamma}U_{isp}$. After $p^t$ is determined, the ISP and the CP
maximize their individual utilities. Here, our study is limited to the cast that
the best strategy $(p^{s*}, q^*, c^*)$ satisfies $p^{s*}>0$, $q^*>0$ and $c^*>0$.
We solve $(p^{s*}, q^*, c^*)$ by letting the derivatives be 0 because
it is not at the boundary. Here, $\frac{U_{cp}}{dc}$ is expressed as
\begin{eqnarray}
\frac{dU_{cp}}{dc} = (1-\gamma) \big((\frac{M^2B^2\bar{v}}{(MB+\bar{v}D)^2} + p^s - p^r)\frac{K}{1+c} -1  \big) = 0. \label{eq:sec4.4.4_cputility_derivative}
\end{eqnarray}
The derivatives $\frac{U_{isp}}{dp^s}$ and $\frac{U_{cp}}{dp^s}$ are
\begin{eqnarray}
\frac{dU_{isp}}{dp^s} &=& \gamma \big(D - \alpha(\frac{M^2B^2\bar{v}}{(MB+\bar{v}D)^2} + p^s - p^r) \big) = 0, \label{eq:sec4.4.5_isputility_derivative1}\\
\frac{dU_{isp}}{dq} &=& \gamma \big(\beta(\frac{M^2B^2\bar{v}}{(MB+\bar{v}D)^2} + p^s - p^r) - 2p^rq\big) = 0.\label{eq:sec4.4.6_isputility_derivative2}
\end{eqnarray}
%
%
%
%
The equations \eqref{eq:sec4.4.4_cputility_derivative} and \eqref{eq:sec4.4.5_isputility_derivative1} give rise to
\begin{eqnarray}
D = \frac{\alpha(1+c)}{K}.\label{eq:sec4.4.7}
\end{eqnarray}
The equations \eqref{eq:sec4.4.5_isputility_derivative1}, \eqref{eq:sec4.4.6_isputility_derivative2} together with the demand function \eqref{eq2.1.2.1_demand3} yield
\begin{eqnarray}
\frac{(4\alpha p^r - \beta^2)D}{2\alpha p^r} - \frac{M^2B^2\bar{v}\alpha}{(MB+\bar{v}D)^2} = D_0^0 - \alpha p^r + K\log(1+c).\label{eq:sec4.4.8}
\end{eqnarray}
The best investment $c^{*}$ is the solution of the fixed point equation composed of \eqref{eq:sec4.4.7} and \eqref{eq:sec4.4.8}
if $c^{*}\geq 0$. 
From \eqref{eq:sec4.4.7} and \eqref{eq:sec4.4.8}, $D$ is strictly increasing w.r.t. $c$ if $c$ and $D$ are positive.
But $D$ increases linearly in \eqref{eq:sec4.4.7} and logarithmically in \eqref{eq:sec4.4.8}.
In order to prove the uniqueness of the best strategy, we
need to show that $D$ computed by \eqref{eq:sec4.4.8} is larger than that computed by \eqref{eq:sec4.4.7}.
When $c$ is 0, the demand $D$ is $\frac{\alpha}{K}$.
We then replace $D$ by $\frac{\alpha}{K}$ in \eqref{eq:sec4.4.8}.
If there has $\frac{(4\alpha p^r - \beta^2)}{2K p^r} - \frac{M^2B^2\bar{v}\alpha}{(MB+\bar{v}\alpha/K)^2} - D_0^0 + \alpha p^r < 0$,
the demand $D$ computed by \eqref{eq:sec4.4.8} is larger than that computed by \eqref{eq:sec4.4.7}. Given the above inequality,
the best strategy $(p^{s*}, q^*, c^*)$ is unique. 
Otherwise, there might have two fixed point solutions, in which one of them bring more utilities to the ISP and the CP than the other.

\section{Evaluation}
\label{sec:evaluation}

We present some numerical results to reveal how the QoS, prices of the ISP and the CP,
as well as their utilities evolve when the price of side payment
changes. The impact of bargaining power on the choice of side payment is also illustrated.

\begin{figure}[!htb]
 \centering
 \subfigure[{\small Prices of the ISP and the CP}]
 {
 \label{fig:price_subs}
 \includegraphics[width=3.2in]{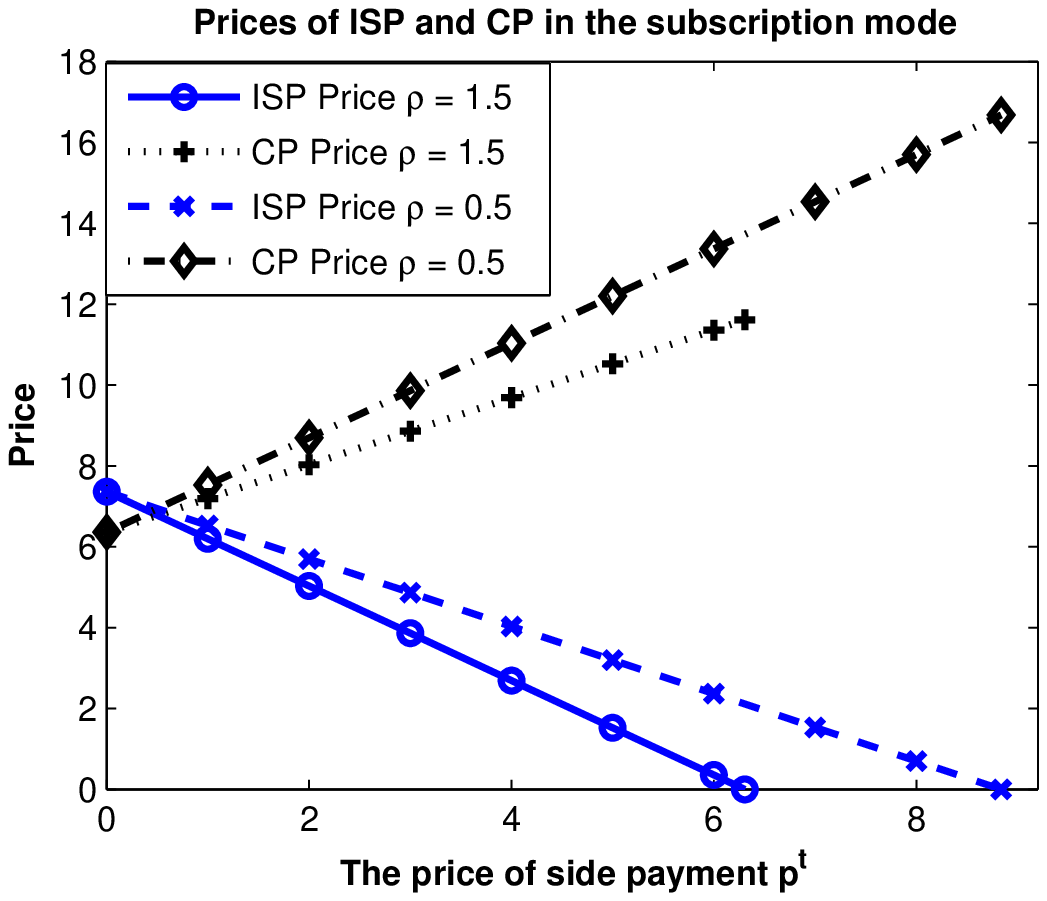}
 }
 \subfigure[{\small Utilities of the ISP and the CP}]
 {
 \label{fig:utility_subs}
 \includegraphics[width=3.2in]{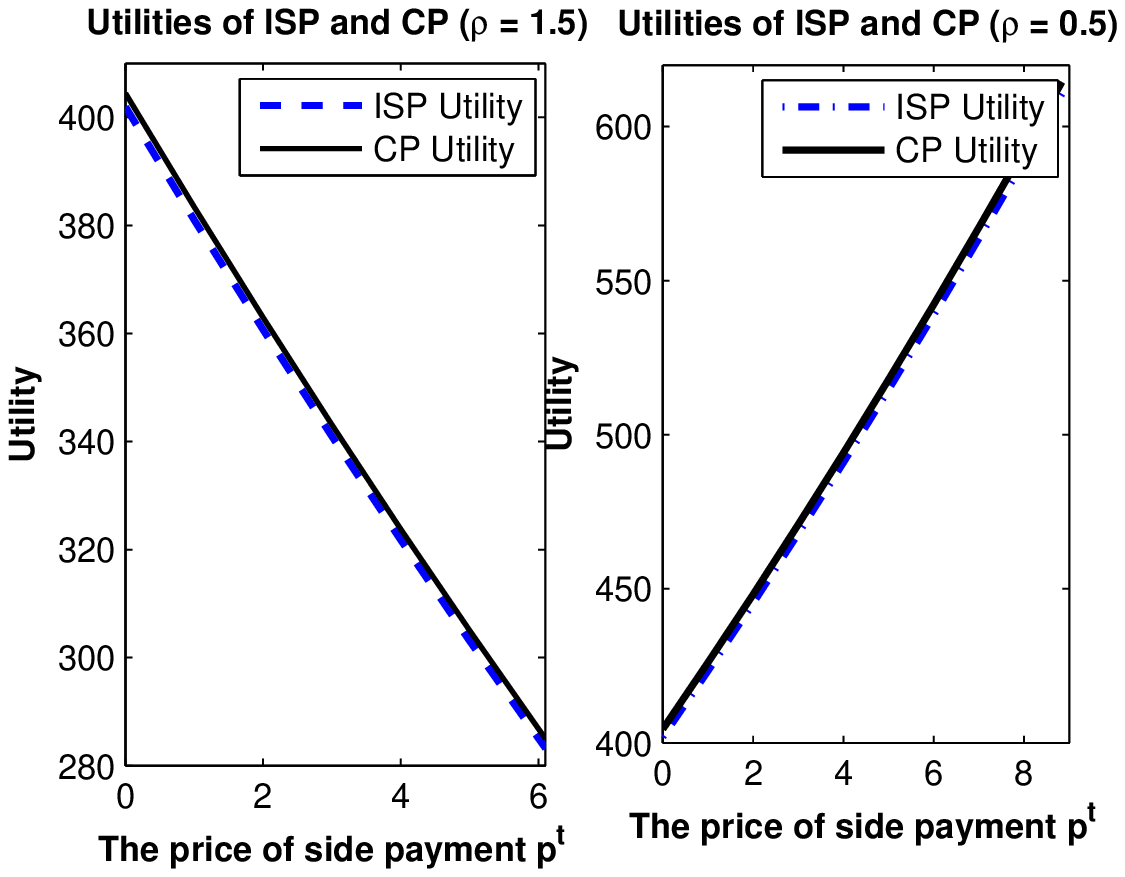}
 }
 \subfigure[{\small The QoS level provided by the ISP}]
 {
 \label{fig:qos_subs}
 \includegraphics[width=3.2in]{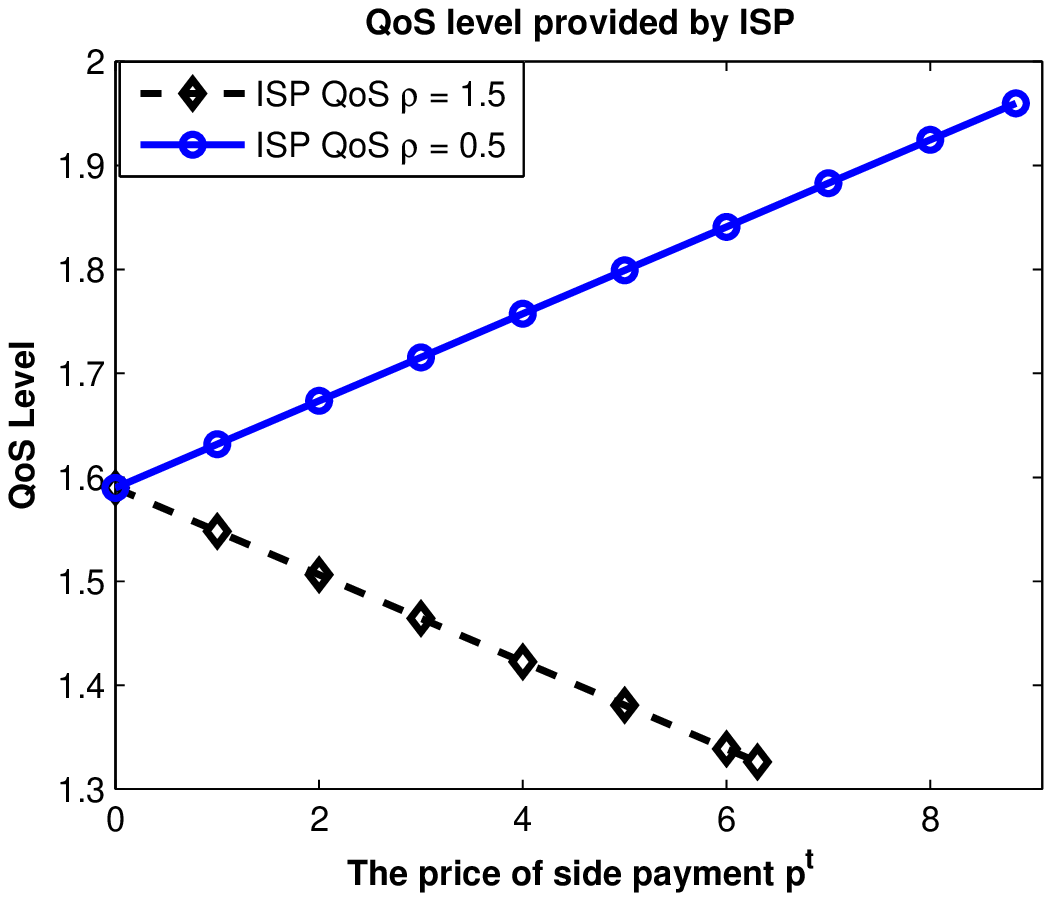}
 }
 \subfigure[{\small Pre-bargaining of $p^t$ : subscription}]
 {
 \label{fig:prebargain_subs}
 \includegraphics[width=3.2in]{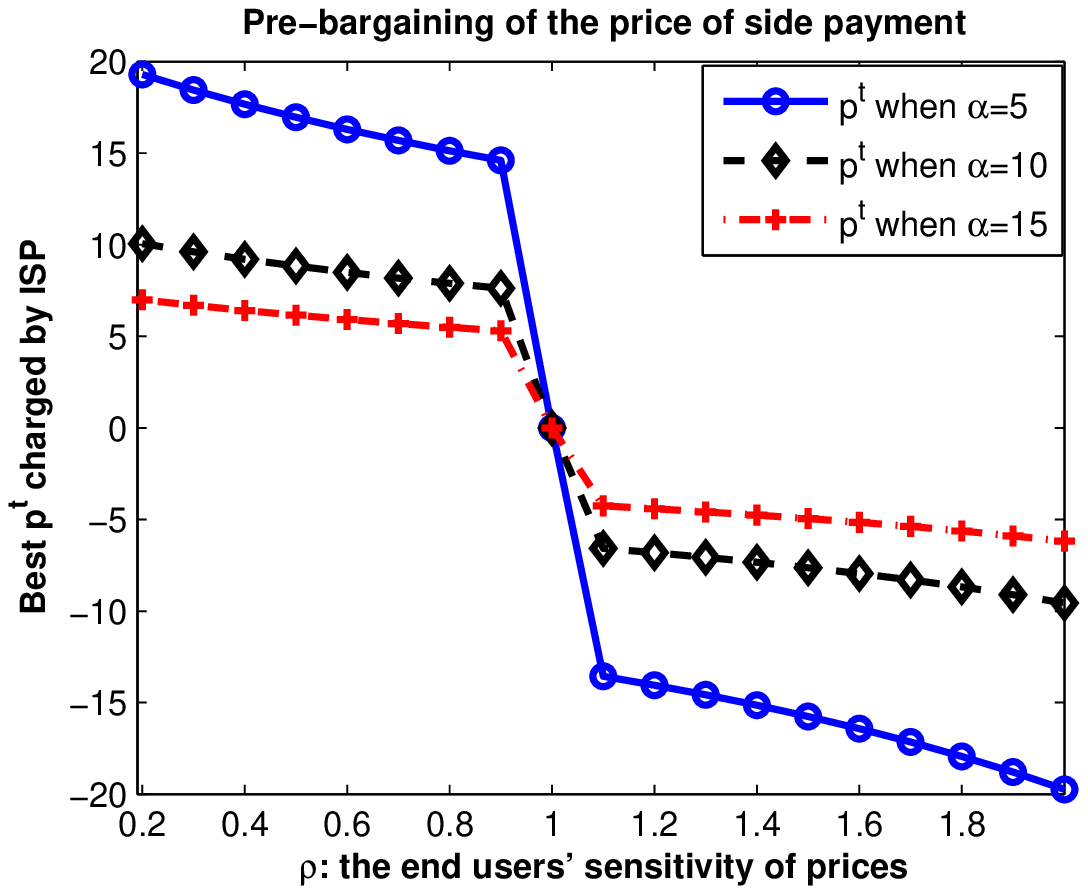}
 }
 \caption{Numerical study for the \emph{subscription} based revenue model}
\end{figure}

\textbf{Subscription Model:}
We consider a networking market where the demand function is given by $D = 200 - 10(p^s + \rho p^c) + 0.5q$.
The operational cost of per-unit of bandwidth is set to $p^r = 1$.
Two situations, $\rho = 0.5$ and $\rho =1.5$, are evaluated. The tax rate $\delta$ is set to 0 for simplicity.
As is analyzed, the side payment benefits the ISP and the CP depending on whether $\rho$ is greater than 1 or not.
Figure \ref{fig:price_subs} shows that $p^c$ increases and $p^s$ decreases when $p^t$ becomes larger.
In figure \ref{fig:utility_subs}, $p^t$ has different impacts on utilities of the ISP and the CP.
When $\rho>1$, end users are more sensitive to the change
of $p^c$ than $p^s$. A positive $p^t$ leads to the increase of $p^c$, causing a tremendous decrease of demand.
Hence, both the ISP and the CP lose revenues w.r.t. a positive $p^t$. Figure \ref{fig:qos_subs}
further shows that a positive $p^t$ yields a better QoS if $\rho < 1$ and a worse QoS if $\rho > 1$.

Next, the ISP and the CP bargain with each other to determine $p^t$. We relax the choice of $p^t$ so that it
can be negative. In the pre-bargaining
game, $p^t$ is \emph{\textbf{independent}} of the bargaining power $\gamma$.
The optimal $p^t$ is obtained when $p^{s*}$ decreases to 0, its lower bound. We evaluate $p^t$ by changing
$\rho$ and $\alpha$ in figure \ref{fig:prebargain_subs}. When $\rho$ increases from 0.2 to 2, $p^t$ decreases until it
becomes negative. A negative $p^t$ means that the ISP needs to transfer revenue to the CP instead. When $\rho = 1$, $p^t$
can be an arbitrary value as long as $p^{s*}$ and $p^{c*}$ are nonnegative. Figure
\ref{fig:prebargain_subs} also shows that a larger $\alpha$ results in a smaller absolute value of $p^t$.

\begin{figure}[!htb]
 \centering
 \subfigure[{\small CP's investment at the equilibrium}]
 {
 \label{fig:invest_adv}
 \includegraphics[width=3.2in]{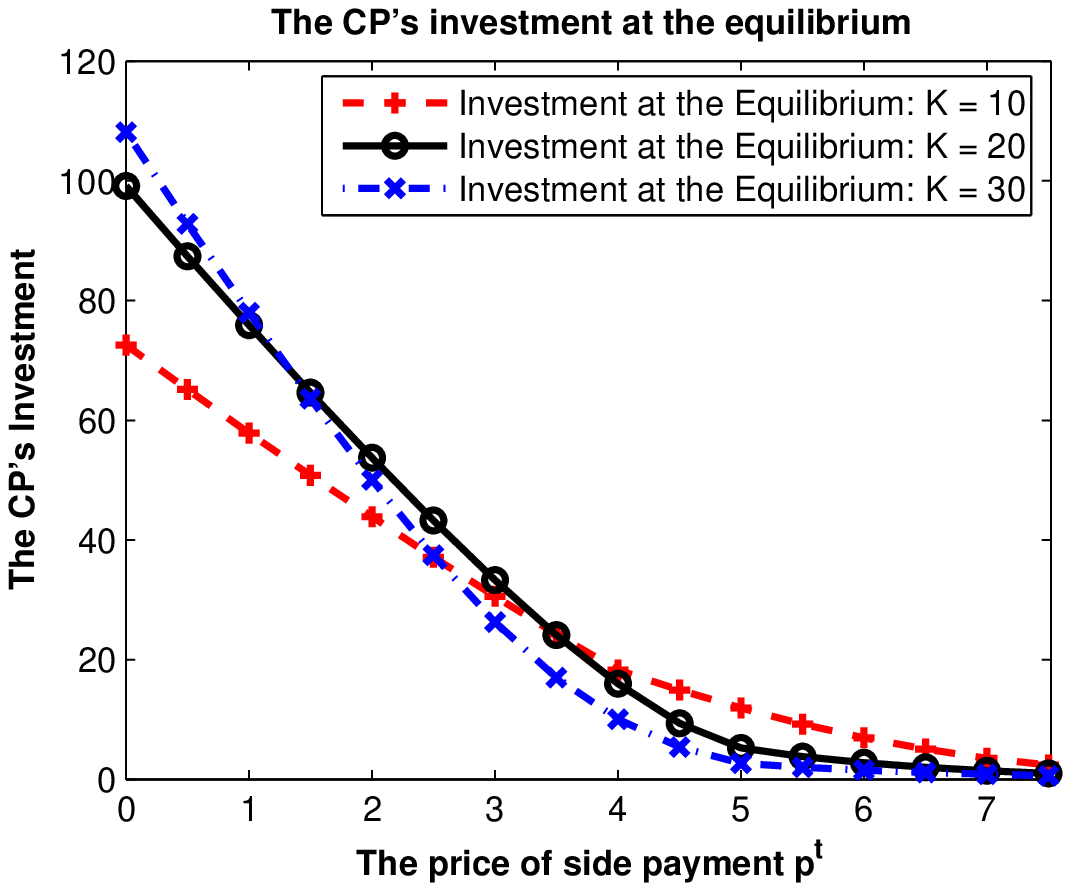}
 }
 \subfigure[{\small The CP's utility at the equilibrium}]
 {
 \label{fig:cputility_adv}
 \includegraphics[width=3.2in]{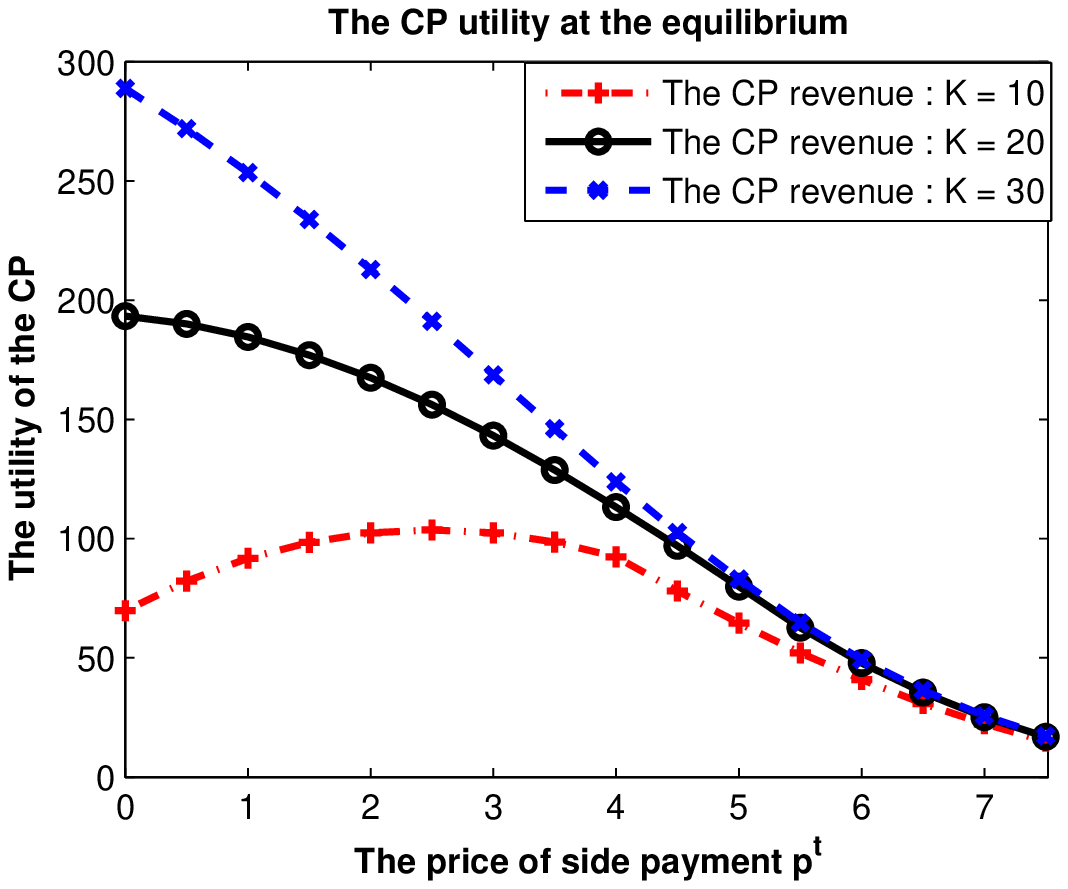}
 }
 \subfigure[{\small The ISP's utility at the equilibrium}]
 {
 \label{fig:isputility_adv}
 \includegraphics[width=3.2in]{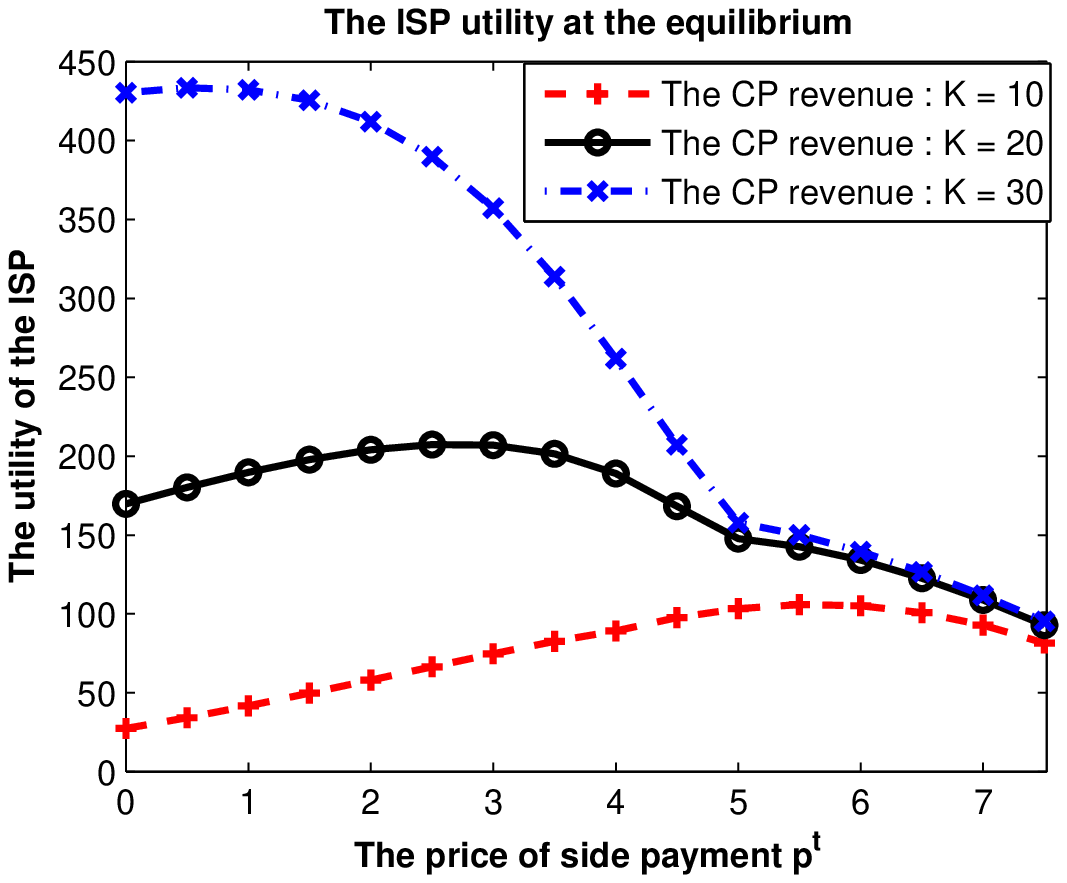}
 }
 \subfigure[{\small Pre-bargaining of $p^t$ : advertisement}]
 {
 \label{fig:prebargain_adv}
 \includegraphics[width=3.2in]{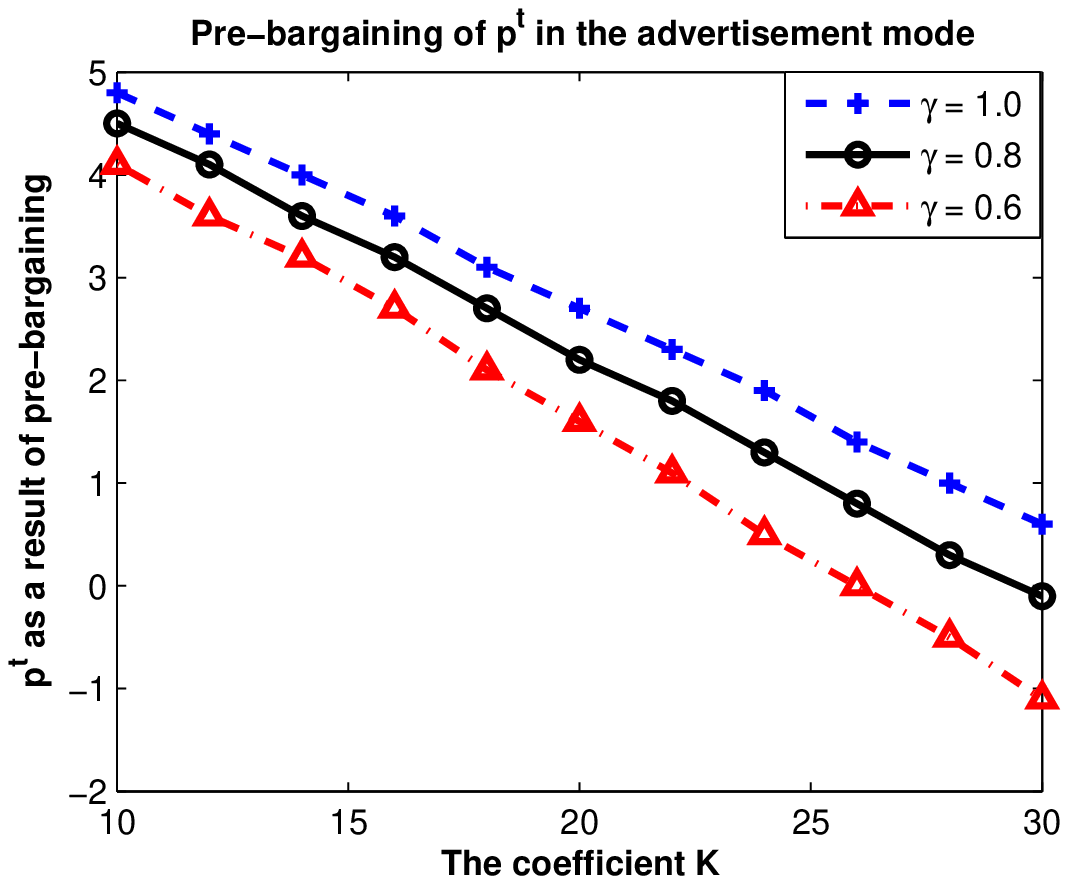}
 }
 \caption{Numerical study for the \emph{advertisement} based revenue model}
\end{figure}

\textbf{Advertisement Model:}
In the advertisement model, we consider the demand function $D = K\log(1+c) - 10p^s + 0.5q$.
The coefficient $K$ reflects the efficiency of the CP's investment to attract end users. The valuation of
each click/browsing follows uniform distribution in the range [0, 10]. The total budget of advertisers
is set to 1000. We conduct two sets of experiments. The first one is to
evaluate the impact of side payment on the best strategies of the ISP and the CP. The second one is to
find the optimal $p^t$ in the pre-bargaining game.
In figure \ref{fig:invest_adv}, the CP's investment is a decreasing function of $p^t$.
When $p^t$ is large enough, $c$ reduces to 0. Figure \ref{fig:cputility_adv}
illustrates the utility of the CP when when $p^t$ and $K$ change.
The CP's utility increases first and then decreases with $K=10$ when $p^t$ increases. For the cases
$K=20$ and $30$, the increase of $p^t$ usually leads to the decrease of revenues.
In figure \ref{fig:isputility_adv}, the utility of the ISP with $K=10$ and $20$ increases first and then decreases when
$p^t$ grows. These curves present important insights
on the interaction between the CP and the ISP. If the contents invested by the CP can bring a large demand,
the side payment is not good for both the ISP and the CP. On the contrary, when the efficiency $K$ is small,
the CP can obtain more utility by paying money to the ISP.

Figure \ref{fig:prebargain_adv} shows the relationship among $K$, $\gamma$ and $p^t$ in the pre-bargaining game.
Different from the subscription model, $p^t$ depends on the bargaining power $\gamma$.
One can observe that $p^t$ is a decreasing function of $K$, and is negative when $K$ is large.
This is to say, the ISP needs to pay money to the CP so that the CP can invest more on its contents.
We also find that the optimal $p^t$ is a decreasing function of $\gamma$.
This is because the ISP is less powerful in negotiating with the CP when $\gamma$ decreases.

\section{Related Work}
\label{sec:related}

There is one particular economic issue that is at the heart of the
conflict over network neutrality. Hahn and Wallsten
\cite{HW} write that net neutrality ``usually means that broadband
service providers charge consumers only once for Internet access, do
not favor one content provider over another, and do not charge
content providers for sending information over broadband lines to end
users.''

This motivates us to study \cite{Eitan1} the implications
of being non-neutral and of charging the content providers.
Using non-cooperative game theoretic tools, we showed that if one
ISP has the power to impose payments on CPs,
not only the end users suffer, but also
the ISP's performance degrades. More
precisely, we show that the only possible equilibrium
(characterized by prices) induces zero demand from the users.
This phenomenon does not occur if the price that the CP is requested
to pay to the ISP is fixed by some regulators.
In \cite{Eitan2} we focus on mechanisms based on the Nash bargaining
(also known as proportional fairness \cite{kelly})
paradigm (which is known in the network engineering context as the
proportional fair assignment). It is the unique way of transferring utilities
that satisfies a well known set of four axioms \cite{nash} related to
fairness. We use a weighted version of this concept that takes
into account the fact that one player may have more weights than
the other one in deciding the amount of side payment.
It is introduced in the context of network neutrality by
\cite{Claudia10}. She analyzes the investment of a CP to duopoly
ISPs for better quality of service.
The bargaining powers of the CP and the ISPs have been shown to
be a knob of investment policies.
In \cite{Eitan3}, we explore the effects of
content-specific (i.e.\ not \emph{application} neutral) pricing,
including multiple CPs providing different types of content.
Also, we consider competition among multiple providers of
the same type and includes different models with consumer stickiness (or loyalty).

Some other recent work includes \cite{IEP07:Hermalin, Economides08, Choi08, Walrand09,asu,CBG}.
Authors in \cite{IEP07:Hermalin} employ the theory
of product-line restriction and find that network neutrality potentially harms the welfare of
small CPs. Economides et. al. in \cite{Economides08}, on the contrary, find that regulated network neutrality
has a better welfare than the two-sided pricing (i.e. pricing both CPs and end users). This
contradictory conclusion is drawn on the assumption that CPs' contents are homogeneous.
In \cite{Choi08}, the authors study the effects of service discrimination on investment incentives
for the ISPs and the CPs, and their implications for social welfare. They show that
net neutrality might not necessarily prohibit the investment of the ISPs, and the discriminatory
regime may weaken the investment incentive of the CPs.
\cite{Walrand09} compares one-sided and two-sided pricing of the ISPs in a network
where the joint investments of the CPs and the ISPs bring revenue from advertisers to the CPs.
Other references that explored the fact that the consumer pays
two entities for one good (for accessing contents through the
Internet) are \cite{asu,CBG}. They study the impact
of this competition on the users as well as on the level
of investment in the infrastructure.

The side payment from the CP to the ISP is expected to be
financed by the income from the users and publicity. Cooperative
games is a well established scientific area that provides us with tools
for designing such mechanisms which, moreover, possess some
fairness properties. In \cite{ToN1:Ma,ToN2:Ma} the Shapley value (which is known to have some
fairness properties \cite{e-winter}) has been used for deciding how
revenues from users should be split between the service and the content
providers.

\section{Conclusion and Future Work}
\label{sec:conclusion}

In this paper, we have studied the two-sided ISP pricing in a non-neutral market
composed of one ISP, one CP, a number of advisers and end users.
We first answer under what situations the \emph{side payment} charged by the ISP is beneficial for the ISP (or the CP).
Then, we study how the price of side payment is determined.
Different from existing work, our models take account of three important features,
the relative price sensitivity, the revenue models, and the QoS provided by the ISP.
These new features give rise to quite different results, enabling us to understand the two-sided pricing comprehensively.
With the subscription model, the relative price sensitivity determines whether the ISP should charge
the side payment from the CP or not. With the advertisement model, the charge of side payment depends on the
ability of the CP's investment to attract the demand.
Our work also has a couple of potential limitations.
For instance, the usage based pricing is usually adopted by wireless ISPs, but not the wireline
ISPs that use the xDSL technology. Sometimes, the CPs might obtain revenue from both end users and the advertisers,
which is not considered here. Our future work would address these limitations.

\end{document}